\documentclass{aa}  

\usepackage{graphicx}
\usepackage{balance}
\usepackage{natbib}
\usepackage{setspace}
\usepackage{txfonts}
\usepackage{placeins}

\newcommand{\kep}{{\it Kepler}}
\newcommand{\prot}{$P_\text{rot}$}
\newcommand{\sph}{$S_\text{\!ph}$}
\newcommand{\teff}{$T_\text{eff}$}

\usepackage{colortbl}
\usepackage{amssymb}
\newcommand{\avsph}{$\langle S_\text{\!ph}\rangle$}
\newcommand{\stdsph}{$\sigma(S_\text{\!ph})$}

\let\OLDthebibliography\thebibliography
\renewcommand\thebibliography[1]{
  \OLDthebibliography{#1}
  \setlength{\parskip}{0pt}
  \setlength{\itemsep}{0pt plus 0.3ex}
}

\definecolor{purple}{RGB}{102,0,204}
\definecolor{blue2}{RGB}{29,95,161}
\definecolor{orange}{RGB}{225,130,0}
\definecolor{green2}{RGB}{29,161,95}

\usepackage{hyperref,multirow}
\hypersetup{colorlinks,citecolor={blue2},linkcolor={blue2},urlcolor={blue2}}  

\begin{document}

   \title{Signature of spin-down stalling in stellar magnetic activity} 
   \subtitle{The case of the open cluster NGC 6811}

   \author{A. R. G. Santos\inst{1,2} 
          \and D. Godoy-Rivera \inst{3,4}
          \and S. Mathur \inst{3,4} 
          \and S. N. Breton \inst{5} 
          \and R. A. Garc\'{i}a \inst{6} 
          \and M. S. Cunha \inst{1}
          }

   \institute{Instituto de Astrof\'isica e Ci\^encias do Espa\c{c}o, Universidade do Porto, CAUP, Rua das Estrelas, PT4150-762 Porto, Portugal \\
   \email{Angela.Santos@astro.up.pt}
   \and Departamento de F\'isica e Astronomia, Faculdade de Ciências, Universidade do Porto, Rua do Campo Alegre 687, PT4169-007 Porto, Portugal
   \and
       Instituto de Astrof\'\i sica de Canarias (IAC), E-38205 La Laguna, Tenerife, Spain
   \and
       Universidad de La Laguna (ULL), Departamento de Astrof\'isica, E-38206 La Laguna, Tenerife, Spain
   \and
       INAF – Osservatorio Astrofisico di Catania, Via S. Sofia, 78, 95123 Catania, Italy
   \and
       Universit\'e Paris-Saclay, Universit\'e Paris Cit\'e, CEA, CNRS, AIM, 91191, Gif-sur-Yvette, France
          }

   \date{\today}
   
   \titlerunning{Magnetic activity of NGC 6811}
   
   \authorrunning{A. R. G. Santos}

  \abstract
   {Stellar rotation and magnetic activity have a complex evolution that reveals multiple regimes. One of the related transitions that is seen in the rotation distribution for main-sequence (MS) solar-like stars has been attributed to core-envelope coupling and the consequent angular-momentum transfer between a fast core and a slow envelope. This feature is known as spin-down stalling and is related to the intermediate-rotation gap seen in field stars.} 
   {Beyond this rotation signature, we search for evidence of it in stellar magnetic activity.}
   {We investigated the magnetic activity of the 1 Gyr old NGC 6811, a \kep-field cluster, and \kep\ MS stars of different ages. The magnetic activity was measured through the photometric magnetic activity proxy, \sph. To characterize the evolution of the magnetic activity for the \kep\ sample, we split it according to the relative rotation and computed the respective activity sequences.}
   {We found the signature of core-envelope coupling in the magnetic activity of NGC 6811 and in the \kep\ MS sample. In NGC 6811, we found enhanced magnetic activity for a range of effective temperatures that remained for significant timescales. In the \kep\ sample, the magnetic activity sequences pile up in two distinct regions: 1) at high activity levels that coincide with stars near the stalling mentioned above, where a behavior inversion is observed (slowly rotating stars have higher activity levels than fast-rotating stars, which is opposite to the overall behavior); and 2) at low activity levels corresponding to slow rotators close to the detection limit, potentially facing a weakening of the magnetic braking.}
   {These results support the recent proposition that the strong shear experienced by stars during the core-envelope coupling phase can cause enhanced activity. This study helps us to shed light on the interplay between rotation, magnetic activity, and their evolution.}


   \maketitle

\section{Introduction}\label{sec:intro}

On the main sequence (MS), rotation and magnetic activity of solar-like stars (low-mass stars with outer convective layers) strongly depend on stellar age \citep[e.g.][]{Wilson1964,Skumanich1972,Mamajek2008,Angus2015, Lorenzo-Oliveira2018}. Solar-like stars start their MS with relatively fast rotation and high activity levels \citep[e.g.][]{Barnes2003,Fritzewski2021b} in the so-called saturated regime, in which magnetic activity is mostly independent of rotation \citep[e.g.][]{Wright2011,Wright2016}. By losing angular momentum through their magnetized winds \citep[magnetic braking; e.g.][]{Weber1967,Kawaler1988,Pinsonneault1989,Gallet2013}, the stars gradually converge to the unsaturated regime and the slow-rotation sequence. In this regime, the magnetic activity depends on the rotation rate, and both decay with stellar age \citep{Wilson1963,Skumanich1972,Soderblom1991}. The spin-down timescales in both regimes depend on the mass \citep[e.g.][]{Barnes2003,Barnes2007,vanSaders2013,Matt2015}. In the saturated regime, higher-mass (typically hotter) solar-like stars converge faster to the rotation sequence, whereas in the unsaturated regime, lower-mass (typically cooler) stars experience a more pronounced rotation evolution. Early-F stars, with an effective temperature higher than $\sim 6250$ K \citep[Kraft break;][]{Kraft1967}, do not spin down significantly in the MS because their shallow convective envelopes cannot harbor efficient dynamos.

The observation of the gradual decay in the rotation rate and magnetic activity by \citet{Skumanich1972} inspired the development of gyrochronology and magneto(gyro)chronology \citep[e.g.][]{Barnes2003,Barnes2007,Garcia2014,Angus2019,Bouma2023,Boyle2023,Mathur2023,Ponte2023,Lu2023,Van-Lane2024,Bouma2024}, which use rotation and/or magnetic activity as proxies of stellar age. These techniques can become particularly important in stars or regimes for which no asteroseismology is available. However, more complete calibration samples and a better understanding of stellar evolution are required.

Space-based photometry has opened new avenues for advancing our understanding by allowing the monitoring of large samples of stars. Namely, the space missions \kep\ \citep{Borucki2010} and \textit{Gaia} \citep{Gaia2016} have contributed significantly to this growth. These two missions collectively provided rotation periods and magnetic activity levels for about half a million stars \citep[e.g.][]{Nielsen2013,McQuillan2014,Ceillier2016,Lanzafame2018,Lanzafame2023,Santos2019a,Santos2021ApJS,Distefano2023,Reinhold2023}. While the \textit{Gaia} sample is larger, the \kep\ and \textit{Gaia} samples are complementary \citep[e.g.][]{Lanzafame2019,Santos2024}: \textit{Gaia} is mostly sensitive to (young) fast-rotating stars, while the \kep\ stars are older and rotate more slowly. Both missions have revealed unexpected details and transitions in the evolution of the rotation and magnetic activity of solar-like stars.

In particular, after MS solar-like stars transitioned from the saturated to the unsaturated regime, their evolution was initially perceived as monotonic, as described above, with a gradually decreasing rotation rate and activity. However, \kep\ revolutionized this perspective by showing that the magnetic unsaturated regime is not continuous, but has a gap at intermediate rotation \citep[e.g.][]{McQuillan2013a,McQuillan2014,Davenport2017,Davenport2018,Mathur2025}. This suggests a phase of fast evolution. The observation of the intermediate-rotation gap in K2 data \citep{Howell2014} and ground-based data \citep[from the Zwicky Transient Facility;][]{Bellm2019} confirmed that the intermediate-rotation gap is a result of stellar evolution and is not restricted to the \kep\ field \citep{Reinhold2020,Gordon2021,Lu2022}. The hypothesis for the causing mechanism that has gathered the most supporting evidence is core-envelope coupling, which promotes the angular-momentum transfer between the envelope that has spun down due to magnetic braking and the fast-rotating core \citep[e.g.][]{Curtis2019,Spada2020,Angus2020,Gordon2021,Lu2022}. This gap is not detected in the fully convective regime \citep[effective temperatures lower than $\sim 3500$ K;][]{Lu2022}. At the opposite end of solar-like stars, the intermediate-rotation gap and consequent bimodal rotation distribution are not observed either for early-F stars, but likely for a different reason. In the region of the parameter space in which stars that rotate faster than the gap were expected to be detected, F stars are almost absent from the \kep\ sample. This might either be an observational bias or be related to the fast evolution of F stars \citep{Mathur2025}.

\kep\ provided another piece of evidence for this transition in the rotation evolution of solar-like stars, particularly through the monitoring of the open cluster NGC~6811, which has an age of $\sim1$ Gyr \citep{Meibom2011b,Curtis2019}. \citet{Curtis2019} found that its rotation sequence deviates from the expected behavior, especially for K dwarfs, which have not spun down over the last 300 Myr. This indicated a spin-down stalling \citep{Curtis2019}. \citet{Cao2023} found the signature of this transition not in the rotation, but in the magnetic activity of the metal-rich $670$ Myr old Praesepe cluster \citep[measured from spectra collected by the Apache Point Observatory for Galactic Evolution Experiment - APOGEE;][]{Majewski2017}. The authors concluded that their observations support the core-envelope coupling scenario because the strong internal shear between the fast core and the slow envelope could explain the observed enhancement in the magnetic activity. 

This transition at intermediate rotation is the focus of this work. We search for its signature in the magnetic activity of the cluster NGC~6811, which was observed by \kep, and in the remainder of the \kep\ single MS stars. Magnetic activity proxies can be inferred from the stellar light curves. Particularly when dark magnetic spots or active regions cross the visible disk of the stars, they modulate the stellar brightness. The amplitude of these brightness variations is related to the active region coverage of the stellar surface \citep[e.g.][]{Basri2013,Mathur2014b,Salabert2017,Santos2023}, which is in turn related to the magnetic activity level of the star. Therefore, the amplitude of the brightness variations due to active regions can be used as a proxy for the stellar magnetic activity. We focus on the magnetic activity proxy \sph\ defined by \citet[][expanding on \citealt{Garcia2010}]{Mathur2014b} as the standard deviation of the flux over segments with a length of five times the stellar rotation period. The \sph\ values are corrected for the photon-shot noise. We note, however, that there are instances where departures from the described relation between \sph\ and active-region coverage can be expected. Specifically, the spatial distribution of active regions in longitude and latitude also impacts the relative amplitude of the brightness variations. In addition, the vantage point depends on the stellar inclination angle, which is mostly unknown. Nevertheless, \sph\ has been shown to be a valid magnetic activity proxy for the Sun and other solar-like stars \citep{Mathur2014b,Mathur2025,Salabert2016a,Salabert2017,Santos2023,Santos2024}.

In Sect.~\ref{sec:sample} we present the samples we analyzed in this work, which contain no potential contaminants, and we focus on single MS stars according to the latest available data. Section~\ref{sec:sph-prot} discusses the discontinuity in the unsaturated regime seen in the activity-rotation relation for \kep\ field stars. The rotation sequence of NGC~6811 and its relation to the rotation distribution of the \kep\ sample is shown in Sect.~\ref{sec:prot}. Section~\ref{sec:results} presents our results, which show the signature of the transition associated with the intermediate-rotation gap (and spin-down stalling) on the magnetic activity of NGC~6811 members and \kep\ field stars. In Sect.~\ref{sec:conclusions} we draw our conclusions.

\section{Target selection: Single main-sequence stars}\label{sec:sample}

First, we selected the reference \kep\ single MS stars with rotation periods (\prot) measured by \kep. We started with the surface rotation sample from \citet{Santos2019a,Santos2021ApJS} and applied the selection criteria from \citet{Godoy-Rivera2025}. This selection discarded evolved stars based on their location in the color-magnitude diagram by empirically determining the locus of the MS, using {\it Gaia} Data Release 3 \citep[DR3;][]{Gaia_DR3} and comparing it with stellar models. Additionally, it discarded candidate binary systems identified as {\it Gaia} variables \citep{Katz2023,Rimoldini2023}, {\it Gaia} nonsingle stars \citep{Binaries_GaiaDR3} and \kep\ eclipsing binaries \citep{Kirk2016}, as well as stars with a renormalized united weighted error larger than $1.2$ \citep[RUWE;][]{Gaia_DR3}. This led to a reference sample of 33,731 single MS solar-like stars with measured \prot\ (out of 55,252). The selection criteria are conservative, but for the purpose of this work, we preferred to disregard all potential pollutants from these samples. For comparison, this selection would leave the sample of \citet{McQuillan2014} with 20,750 stars (out of 34,030). We adopted the effective temperature (\teff) from \citet{Berger2020} or, when this was not available, from \citet{Mathur2017}.

Second, for the 1 Gyr-old NGC 6811, we adopted the cluster membership by \citet[][classification of ``probable member'']{Godoy-Rivera2021_cluster} and the values for \prot\ and the average photometric magnetic activity proxy (\avsph) retrieved by \citet{Santos2019a,Santos2021ApJS}. We further applied the selection criteria described above to remove potential binary systems and evolved stars. Five close-in binary candidates flagged in \citet{Santos2019a,Santos2021ApJS} were also removed. This led to a sample of 141 single MS members of the open cluster NGC 6811 with measured \prot\ and \avsph.

\section{Activity-rotation relation for \textit{Kepler} field stars}\label{sec:sph-prot}

The \prot\ distribution of \kep\ MS solar-like stars is known to be bimodal \citep[e.g][]{McQuillan2013a,McQuillan2014,Davenport2018}. At cool temperatures, the two sequences or stellar populations are separated by a rotation gap. In addition to the \kep\ sample, this feature has also been observed in the different K2 campaigns \citep{Reinhold2020,Gordon2021} and in ground-based data \citep{Lu2022}. The observation of this gap at intermediate \prot\ in different surveys with distinct targets suggests that the bimodality is related to stellar evolution. 

\begin{figure}[b]
    \centering
    \includegraphics[width=\hsize]{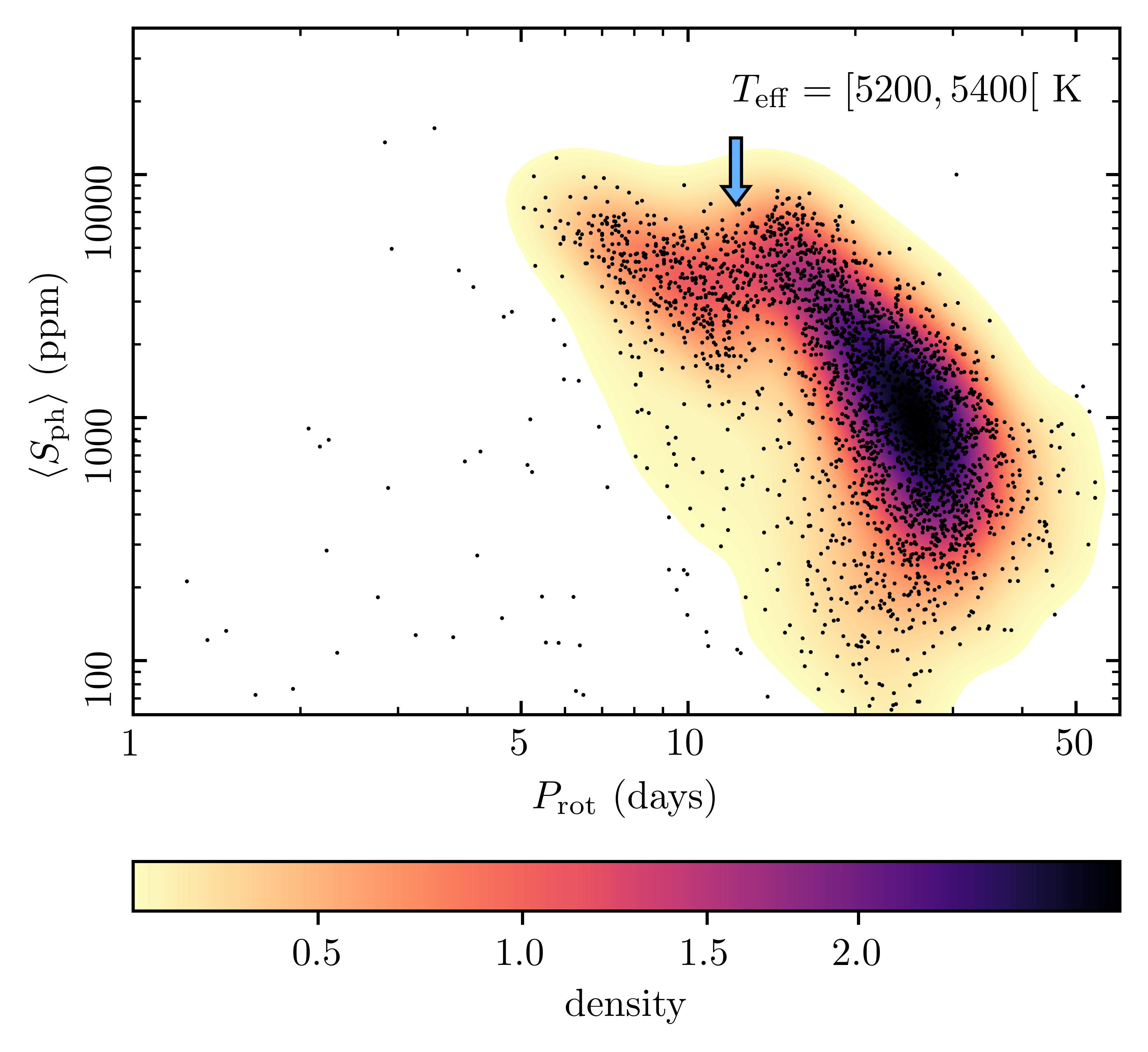}
    \caption{Activity-rotation diagram for \kep\ single dwarfs with temperatures within 5200 and 5400 K, corresponding to the interval in which the stars belonging to NGC 6811 seem to cross the intermediate-\prot\ gap. The black dots correspond to the individual stars, and the colored regions indicate the density of stars, as computed by the kernel density estimation. The blue arrow marks the local minimum in the \sph-\prot\ relation associated with the intermediate-\prot\ gap.}
    \label{fig:sph_prot}
\end{figure}

The bimodality can also be detected in the activity-rotation relation, which breaks into two sequences. A clear distinction between the two sequences depends on the \teff\ range in analysis, however \citep[e.g.][]{Santos2021ApJS,Santos2023,Mathur2025}. This relation is shown in Fig.~\ref{fig:sph_prot} for a subsample of the \kep\ sample within 5200 and 5400 K (shaded region in the inset in Fig.~\ref{fig:prot_teff}). \kep\ provides limited insight into the saturated regime, which corresponds to $P_\text{rot}\lesssim 5$ days in Fig.~\ref{fig:sph_prot} \citep[see discussion in][]{Santos2024}. For $P_\text{rot}\gtrsim 5$ days, two sequences of stars can be identified, separated by a lower-density region that corresponds to the intermediate-\prot\ gap (indicated by the blue arrow in Fig.~\ref{fig:sph_prot} and the dot-dashed line in Fig.~\ref{fig:prot_teff}). Hereafter, we use the terms fast-rotating population and slowly rotating population when we refer to stars that rotate faster and more slowly than the gap, respectively. 

While stars generally become less active as they spin down, some stars in the fast-rotating population show lower magnetic activity levels than the stars in the slowly rotating population around the transition associated with the rotation gap (Fig.~\ref{fig:sph_prot}). This behavior was previously identified and used to estimate the location of the intermediate-\prot\ gap \citep{Reinhold2020,Santos2023}, namely through the local minimum (indicated by the blue arrow).

\section{Rotation sequence of NGC 6811}\label{sec:prot}

NGC 6811 is an open cluster with an age of around 1 Gyr \citep[e.g.][]{Meibom2011b}. By investigating the rotation sequence of NGC 6811, \citet{Curtis2019} observed that the rotation periods of NGC 6811 K dwarfs (within the range 4500 K $<T_\text{eff}< 5000$ K) did not evolve significantly in comparison with the rotation sequence of younger clusters, in particular, the $\sim 670$ Myr old Praesepe. The authors concluded that the K-type stars in this 1 Gyr old cluster experience a stalling in the spin down.

\begin{figure}[b]
    \centering
    \includegraphics[width=\hsize]{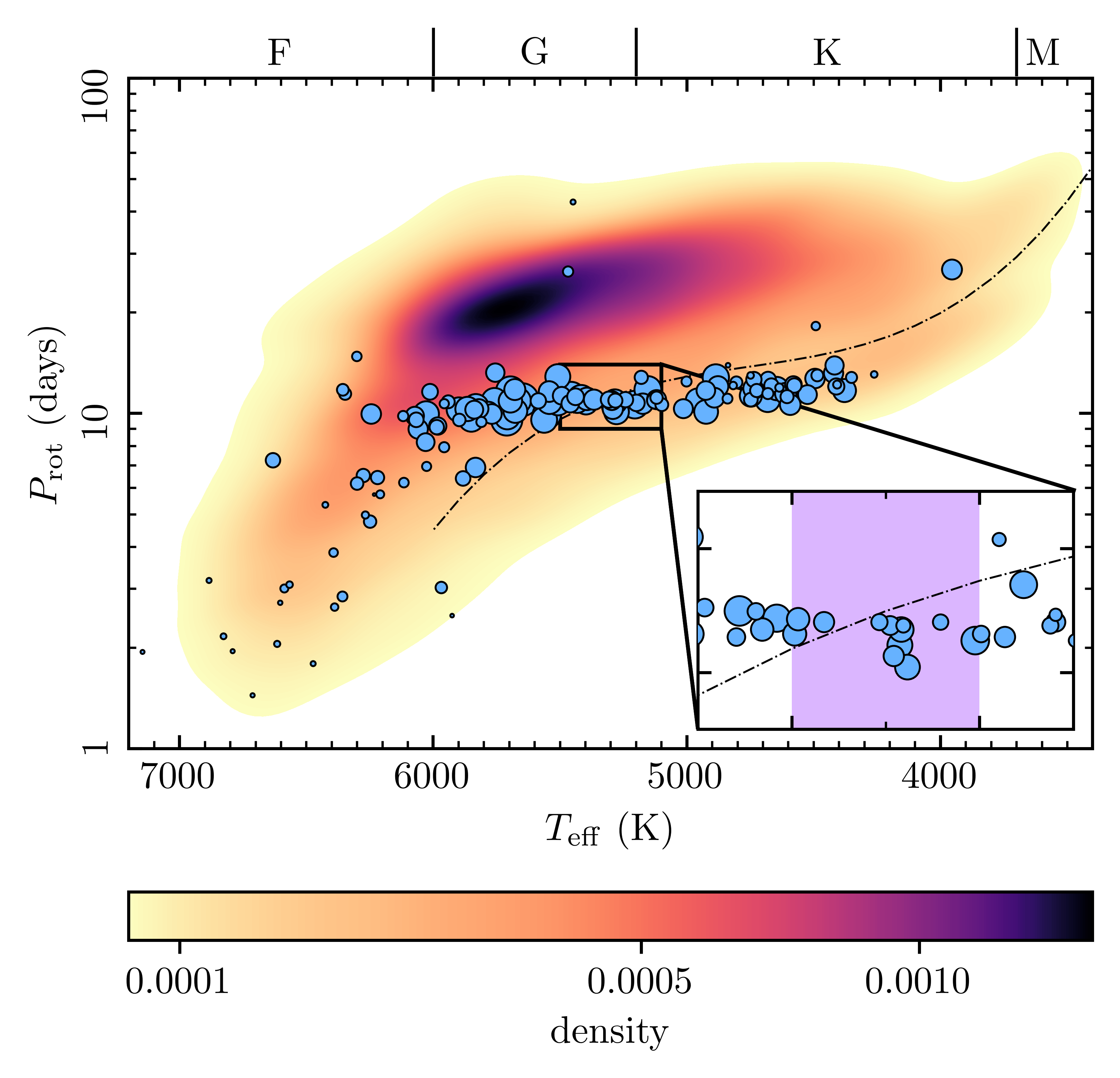}
    \caption{Rotation period as a function of the effective temperature for \kep\ MS single stars. The color-code indicates the density of stars. The dot-dashed line marks the intermediate-\prot\ gap. The blue circles highlight the stars belonging to NGC 6811. The symbol size indicates the magnetic activity level inferred through \avsph: the larger the symbol, the stronger the magnetic activity. The subpanel zooms into the stars of NGC 6811 within the temperature range in which they cross the gap. The shaded purple region corresponds to the 5200 K $<T_\text{eff}<5400$ K interval considered in Fig.~\ref{fig:sph_prot}.}
    \label{fig:prot_teff}
\end{figure}

Figure~\ref{fig:prot_teff} compares the NGC 6811 rotation sequence with the rotation distribution of the reference sample of \kep\ single MS solar-like stars. The dot-dashed line marks the intermediate-\prot\ gap computed using the clean sample found in Sect.~\ref{sec:sample} \citep[following the procedure in][see also our Appendix~\ref{sec:app_gap}]{Santos2023}. Some stars seem to be outliers in the rotation sequence, namely a few slow and fast rotators relative to the cluster rotation sequence. This suggests that the sample may still contain some contaminants that passed the selection criteria summarized in Sect.~\ref{sec:sample}. It is worth mentioning that although rare, these rotational outliers are indeed present in the \prot\ sequence of most clusters \citep[e.g.][]{Godoy-Rivera2021_cluster}.

As shown in Fig.~\ref{fig:prot_teff}, the rotation sequence of NGC 6811 crosses the intermediate-rotation gap. Therefore, these two observations (intermediate-\prot\ gap and spin-down stalling) are thought to be related and correspond to the same transition in the stellar evolution. If this is the case, NGC 6811 members with \prot\ longer than the intermediate-\prot\ gap have already completed the transition.

\section{Results}\label{sec:results}

We searched in the magnetic activity of the selected stars for the signature of the transition associated with the intermediate-\prot\ gap and the observed stalling of the spin-down. We first investigated the activity of the NGC 6811 members (Sects.~\ref{sec:sph_NGC6811} and \ref{sec:bias}). The remainder \kep\ single MS stars are analyzed in Sect.~\ref{sec:kepler}.

\subsection{Activity-temperature relation for NGC 6811}\label{sec:sph_NGC6811}

The stellar Rossby number (Ro), that is, the ratio of \prot\ and the convective turnover time ($\tau_\text{C}$), is related to the efficiency of the dynamo. It generally scales with magnetic activity \citep[e.g.][]{Noyes1984b,Soderblom1993,Wright2011,Wright2016,See2021}: Low (high) Ro values are associated with strong (weak) activity. At a fixed age on the main sequence, $\tau_\text{C}$ and \prot\ both decrease with increasing \teff\ \citep[e.g.][]{Noyes1984b,vanSaders2013,Matt2015,Fritzewski2021b,Corsaro2021}. This results from the shallow convection zone in hot solar-like stars that leads to shorter $\tau_\text{C}$ and to a less efficient spin-down, allowing stars to remain relatively fast rotators. However, the decrease in $\tau_\text{C}$ as \teff\ increases is more drastic than that of \prot\, and hotter stars have higher Ro than cooler stars \citep[e.g.][]{Kim1996,Landin2010,Breton2022}. At a fixed age, the dynamo efficiency is therefore expected to decrease with increasing \teff, corresponding to decreasing depth of the convection zone. This monotonic behavior can be seen in relatively young clusters, such as for the stars in the unsaturated regime of the 300 Myr old NGC 3532 \citep[][e.g. seen in their fig. 7]{Fritzewski2021b}. By comparison, the activity sequence of NGC 6811 is expected to exhibit a similar behavior.

\begin{figure}[h]
    \centering
    \includegraphics[width=\hsize]{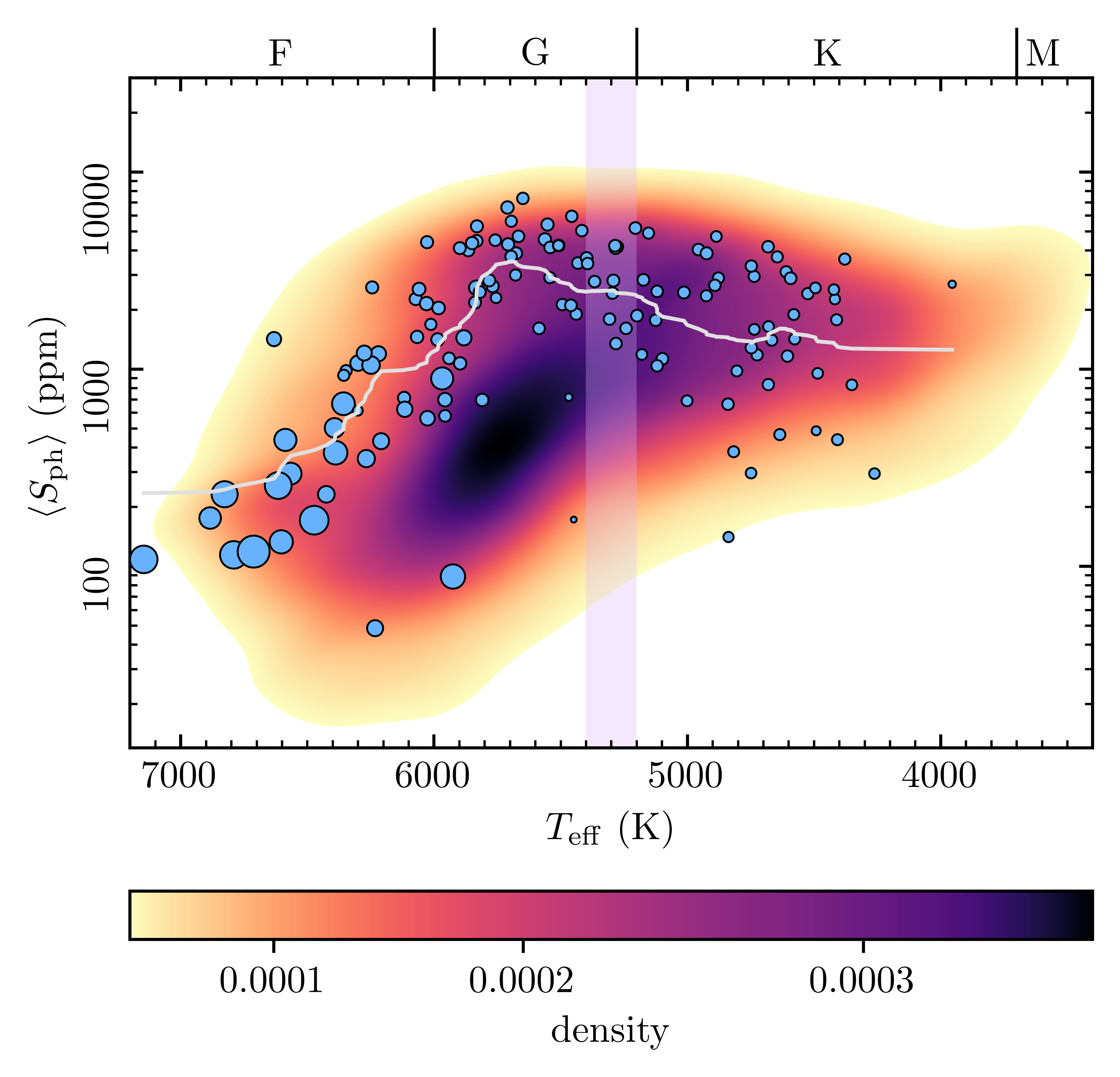}
    \caption{Similar to Fig.~\ref{fig:prot_teff}, but for the average photometric magnetic activity. NGC 6811 members are highlighted by the blue circles, whose size indicates the rotation rate: the larger the symbol, the faster the stellar rotation. The gray line corresponds to the smoothed \avsph\ for the cluster. The semitransparent shaded region marks the \teff\ range highlighted in the inset of Fig.~\ref{fig:prot_teff}. The density map represents the reference \kep\ sample of field stars.}
    \label{fig:sph_teff}
\end{figure}

In Fig. \ref{fig:sph_teff} we investigate the magnetic activity of NGC 6811 members, which shows a different behavior than the monotonic behavior described in the paragraph above. We find that the magnetic activity increases with \teff\ within $\sim 5000$ and $\sim5800$ K. In particular, we encounter an activity enhancement for stars that have already transitioned, that is, those that crossed the intermediate-\prot\ gap ($T_\text{eff} \gtrsim 5300$ K). This observation is consistent with the behavior seen in the activity-rotation relation (Fig.~\ref{fig:sph_prot}) for the reference \kep\ sample, where stars after the transition have higher activity levels than stars before the transition \citep[see the discussion in][]{Mathur2025}.

\begin{figure*}[h]
    \centering
    \includegraphics[width=\hsize]{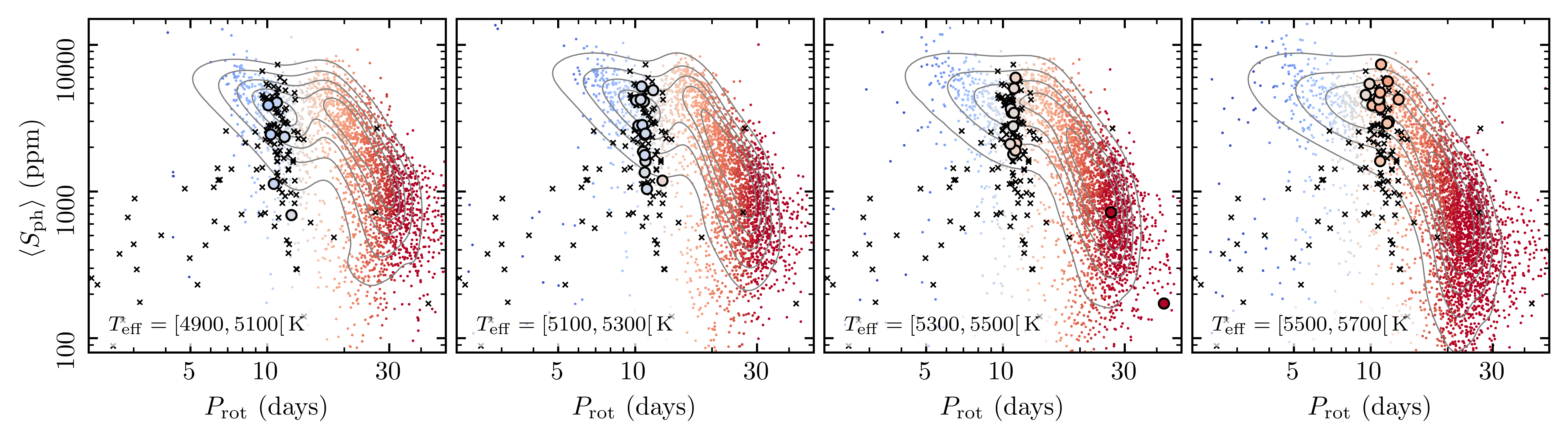}
    \caption{Activity-rotation diagram for the NGC~6811 members (circles) in four \teff\ ranges around the transition associated with the intermediate-\prot\ gap. To better identify the gradual \sph\ shift, the crosses represent NGC 6811 members, but of any \teff. The reference \kep\ single MS sample is shown by the dots and density contours in gray. The color-code for the circles and dots is related to the distance to the intermediate-\prot\ gap, which corresponds to the difference between the logarithm of a given star's \prot\ minus the logarithm of the gap location (truncated at $\delta \log \pm0.4$ for illustration purposes). The lightest colors correspond to the stars closest to the gap.}
    \label{fig:sph_prot_ngc6811}
\end{figure*}

Figure~\ref{fig:sph_prot_ngc6811} shows the activity-rotation diagram of NGC~6811 for different \teff\ ranges with the reference \kep\ sample for comparison. A gradual shift toward higher activity levels for the cluster members can be observed with increasing \teff. In contrast to the reference \kep\ sample, which comprises stars of different ages and chemical composition, the transition signature is seen here in the magnetic activity of a coeval sample. This suggests that the spin-down stalling and the associated intermediate-\prot\ gap leads to a magnetic activity enhancement (from an average \sph\ of 2026.4 ppm at $\sim 5000$ K to 4015.1 ppm at $\sim 5600$ K; the standard deviations are 1248.7 and 1352.5 ppm, respectively). For stars hotter than $\sim 5800$ K, which have transitioned longer ago, the standard activity-\teff\ relation is recovered (e.g., that seen for stars in NGC~3532), with a gradually decreasing activity with \teff\ (Fig.~\ref{fig:sph_teff}).

\subsection{Possible sources of bias}\label{sec:bias}

In the appendices \ref{sec:app_Sphvariation}-\ref{sec:low_activity}, we discuss and explore possible biases to the results above in detail: After the transition associated with spin-down stalling and intermediate-\prot\ gap, stars exhibit enhanced activity levels. In this section, we briefly summarize our conclusions regarding those biases.

Magnetic activity varies at different timescales. The most notable timescale would be the spot cycle, but magnetic activity varies on both shorter and longer timescales \citep[e.g.][]{Olah2009,Bazilevskaya2014,Egeland2015}. The question is whether the enhanced \avsph\ values found in NGC~6811 can be explained as a result of this variability. \citet{Santos2023} investigated the \sph\ variation (\stdsph) over the four years of \kep\ observations. Similar to what was found for chromospheric emission \citep[e.g.][]{Wilson1978,Radick1998,Radick2018,EBrown2022}, the authors found that stars with higher mean photometric activity levels (\avsph) exhibit more significant temporal variations (\stdsph) than stars that are weakly active on average ($\log \sigma(S_\text{ph})\propto 0.842\times\log\langle S_\text{ph}\rangle$; see Appendix~\ref{sec:app_Sphvariation}). The stars in the open cluster NGC~6811 are no exception to this relation (Fig.~\ref{fig:stdsph}). In the case of NGC~6811, the stars that recently transitioned are among those with the highest \stdsph. Nevertheless, we found no clear relation between the distance to the intermediate-\prot\ gap and \stdsph. This suggests that the \sph\ variability does not preferentially affect the \avsph\ measurements of stars near the gap. We found the same trend in the \sph-\teff\ relation regardless of whether we adopted the average, the minimum, or the maximum \sph\ values observed over the four-year \kep\ observations (Fig.~\ref{fig:sigsph_teff}).

Another possible bias could arise from the uncertainty in the location of the intermediate-\prot\ gap and, consequently, in assessing which NGC~6811 stars could have already transitioned. By resampling the \kep\ reference sample according to the \prot\ and \teff\ uncertainties, we inferred the uncertainty associated with the location of the intermediate-\prot\ gap (Appendix~\ref{sec:app_gap}). We found that the location of the intermediate-\prot\ gap is not significantly affected by the uncertainties on \teff\ and \prot\, and thus, our results remain unchanged.

\subsection{Activity-temperature relation for \textit{Kepler} field stars}\label{sec:kepler}

For the \kep\ single MS solar-like sample, the magnetic activity generally decreases with increasing effective temperature (Fig.~\ref{fig:sph_teff}). As discussed above, this observation is consistent with the expected dependence of the dynamo efficiency on \teff, even though stars in this sample have different ages. Nevertheless, there is some structure in the observed relation. Namely, the \kep\ M dwarfs ($T_\text{eff}\!<\!3700$ K) exhibit high activity levels. While the \kep\ M-dwarf sample is very limited, this behavior is consistent with M dwarfs being found to harbor significantly stronger magnetic fields and higher flaring rates than hotter stars \citep[e.g.][]{Johns-Krull1996,Kochukhov2017,Lin2019,Feinstein2020}. Then, there is a \avsph\ slight decrease up to $\sim 4000$ K. While the lower envelope of the \avsph\ distribution continues to decrease, the upper envelope increases between temperatures of $\sim 4000$ and $\sim 5500$ K, before it decreases toward the F stars ($T_\text{eff}\ge6000$ K). 

In Fig.~\ref{fig:sph_teff}, two overdensities can be identified for the \kep\ main-sequence single stars. One overdensity is consistent with the general trend of a decreasing magnetic activity with \teff, and the second overdensity seems to branch out toward higher activity levels.

To search in the \kep\ sample for the activity signature of the spin-down stalling that we identified in Sect.~\ref{sec:sph_NGC6811}, we split the sample according to the distance from the intermediate-\prot\ gap, namely the difference between the logarithms of the respective \prot\ and the gap location, $\delta\log P_\text{rot}$ (see Fig.~\ref{fig:prot_teff_gapcolor}). In other words, we split the reference \kep\ sample into different $\delta \log P_\text{rot}$ bins. We must note that this does not correspond to splitting the sample into coeval subsamples (as is the case for NGC~6811). For those with \prot\ longer than the gap, this selection corresponds to a rough estimate of the time that elapsed since the stars faced the transition associated with the gap. However, we must note that the spin-down timescale depends on the mass \citep{vanSaders2013,Matt2015}, and thus, this selection is a reasonable but approximate calculation.
We also note that each \prot\ value has an associated uncertainty, which is $9.7\%$ on average for this sample. This partially accounts for differential rotation: For the solar case, \prot\ varies by $\sim10\%$ within the active latitudes, for instance. Differential rotation can lead to a distorted and broad rotation peak or to the appearance of multiple rotation peaks in the reference rotation diagnostic. When the amplitude of the differential rotation is relatively high, then it might not be fully retained in the uncertainty \citep[see e.g. the example of KIC 3733735\footnote{Also known in the asteroseismic community as Shere Khan.}, which seems to harbor two distinct active bands in][]{Mathur2014}. There is also some ambiguity in determining the location of the gap and the uncertainty associated with the effective temperature (see Appendix~\ref{sec:app_gap}). Metallicity also affects the rotation and magnetic activity evolution of stars \citep[][]{Karoff2018,Amard2020,Amard2020b,See2021,Santos2023,Mathur2025}. With all of the above in mind, we divided our sample mainly for comparison purposes. 

\begin{figure}[t]
    \centering
    \includegraphics[width=\hsize]{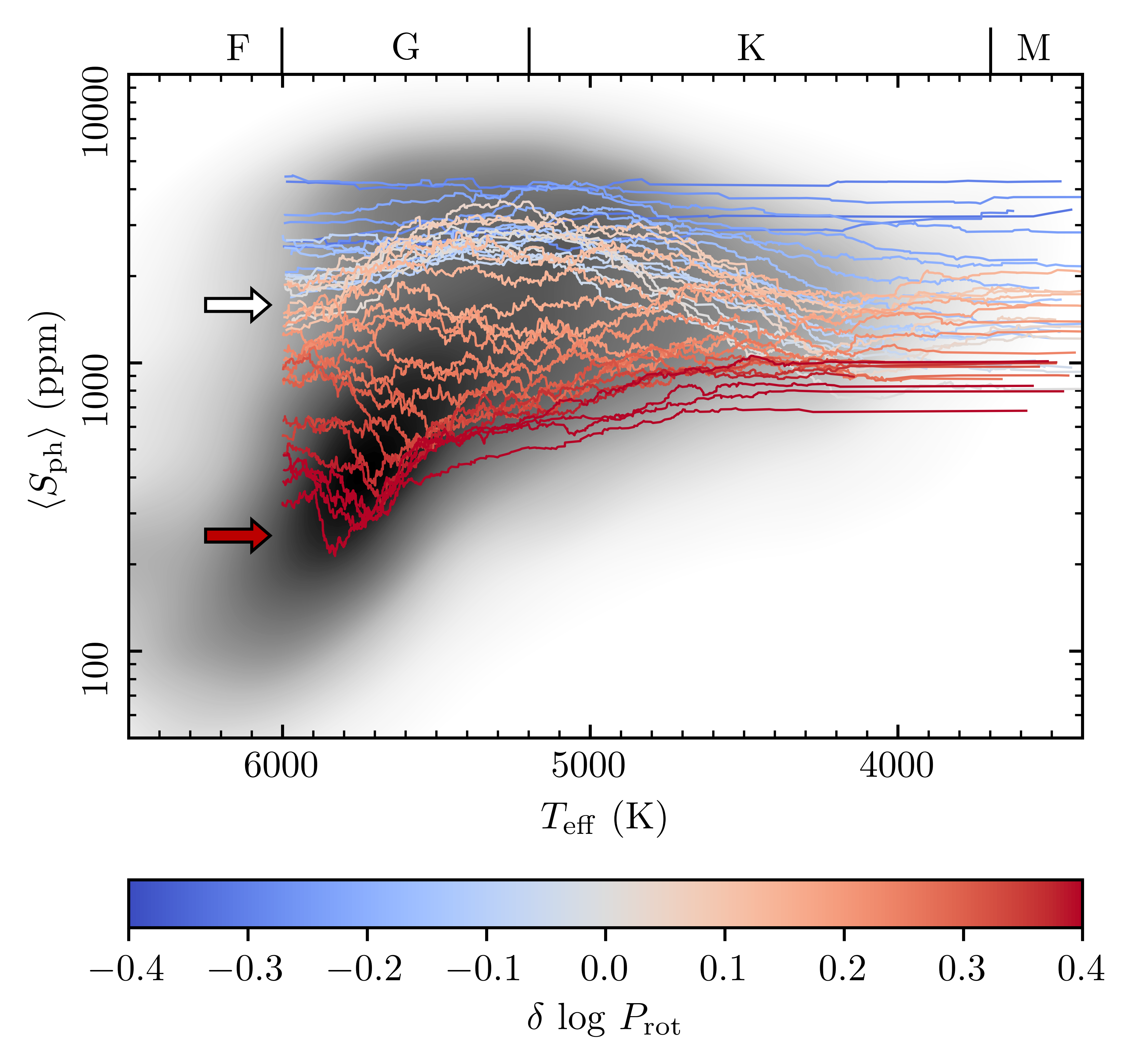}
    \caption{Magnetic activity sequences according to the distance to the intermediate-\prot\ gap, indicated by the color-code. Lighter colors correspond to stars close to the gap. For better visualization, the axis scales differ from those in other figures. To guide the eye, the arrows approximately indicate the activity-sequence overdensities at 6000 K.}
    \label{fig:sph-teff-kep}
\end{figure}

For each subsample (i.e., a given $\delta \log P_\text{rot}$ bin; bin size of 0.02), we obtained the smoothed \sph-\teff\ relation by applying a uniform filter. The resulting curves are shown in Fig.~\ref{fig:sph-teff-kep}, where we kept the full density plot in gray for reference. Curves built with fewer than 50 data points (small samples) are not shown in the figure. The curves stop at 6000 K because the intermediate-\prot\ gap and respective \prot\ bimodality are not found for hotter temperatures (see Sect.~\ref{sec:intro}). The magnetic activity sequences are concentrated around the two overdensities that we identified in the density map (highlighted in Fig.~\ref{fig:sph-teff-kep} with the arrows, but also visible in Fig.~\ref{fig:sph_teff}). The overlapping activity sequences shown in dark red correspond to very slow rotators (red arrow). Pile-ups of \kep\ stars were identified at slow rotation \citep[both in \prot\ and $\text{Ro}$;][]{David2022,Metcalfe2023}. We identified a pile-up in terms of magnetic activity that might be related to the \prot\ pile-up and is consistent with the \prot\ plateau found for old \kep\ stars \citep{Mathur2023}. In Appendix~\ref{sec:low_activity} we investigate whether this behavior can arise from limitations in detecting the low-amplitude signals of weakly active stars. When we removed the faintest targets, whose light curves are typically noisy, we still found this pile-up of activity sequences.

The other pile-up of magnetic-activity sequences occurs around the intermediate-\prot\ gap (light color shades in Fig.~\ref{fig:sph-teff-kep}; indicated by the white arrow). Some of these curves cross each other: For example, some of the curves for stars with $\delta \log P_\text{rot}>0$ (light red) are located at higher \sph\ than those of stars with \prot\ slightly shorter than the gap (light blue). This is consistent with the nonmonotonic behavior seen in the activity-rotation relation (Figs.~\ref{fig:sph_prot} and \ref{fig:sph_prot_ngc6811}).

\section{Discussion and conclusions}\label{sec:conclusions}

It has become increasingly evident that rotation and magnetic activity evolution on the main sequence is complex and deviates from the monotonic behavior described by the so-called Skumanich laws \citep{Skumanich1972}. A significant part of this evidence has been collected based on \kep\ data \citep[see recent review by][]{Santos2024}.

The focus of this work was the transition within the unsaturated regime that is associated with the intermediate-rotation gap. This gap corresponds to a lower-density region in the \prot\ distribution that is bimodal \citep[e.g.][]{McQuillan2013a,McQuillan2014}. This transition can also be explained as a stalling in the rotation evolution, which depends on the mass \citep{Spada2020}. NGC~6811 has been a banner in this regard \citep{Meibom2011b,Curtis2019}, as its rotation sequence reveals a stalling in the rotation evolution for K dwarfs.

We found evidence of the signature of this transition in the magnetic activity of NGC~6811 members, as well as further evidence in the \kep\ MS rotational sample. The latter is in agreement with the findings in \citet{See2021} and \citet{Mathur2025}.
The rotational sequence of NGC~6811 crosses the intermediate-\prot\ gap ($T_\text{eff}\sim 5300$ K). This suggests that some of its members have already undergone the associated transition, while the remaining cooler members have not. We found an enhancement in magnetic activity around the temperatures of this crossing that peaked at significantly higher temperatures. The wide range of temperatures (between $\sim 5000$ and $\sim 5800$ K) of enhanced activity indicates that the effect from the associated transition lasts for a significant time in terms of stellar evolution. The signature also starts at lower temperatures than the crossing, suggesting that the transition already occurs before the intermediate-\prot\ gap, which is a phase of rapid evolution.

Investigating the reference \kep\ MS sample, which is not coeval (contrary to NGC~6811), we searched for evidence for this magnetic activity enhancement. We found two overdensities in the activity-\teff\ relation. The first of these corresponds to very low \avsph\ values and stars with \prot\ significantly longer than the intermediate-\prot\ gap. This overdensity might be related to the weakening in the magnetic braking and the pile-ups found in terms of rotation \citep[instead of magnetic activity;][]{David2022,Metcalfe2023}. The second overdensity corresponds to stars with \prot\ close to the intermediate-\prot\ gap. We found that the activity sequences intercept each other, and sequences corresponding to later stages of stellar evolution show higher activity levels than earlier sequences. Therefore, we conclude that the reference \kep\ sample also shows evidence of the signature of spin-down stalling in the magnetic activity. This agrees with the results for NGC~6811. 

The evidence for magnetic activity enhancement found in this work is consistent with the signature identified for the $\sim 670$ Myr old Praesepe cluster by \citet{Cao2023}. The authors found unexpectedly high activity levels for stars with \teff\ within $\sim4000$ and $\sim4400$ K. At first glance, this would be incompatible with the parameter space of our results for NGC~6811. Nevertheless, in addition to the $\sim300$ Myr difference in the cluster ages, there is another parameter, metallicity, that has a significant impact on stellar evolution, in particular on magnetic activity and rotation, and metallicity might cause this dissimilarity. By changing the opacity, metallicity affects the depth of the convection zone, with metal-rich stars having deeper convection envelopes than their counterparts \citep[e.g.][]{vanSaders2012}. Consequently, magnetic activity is stronger in metal-rich stars \citep[e.g.][]{Karoff2018,See2021,See2023,Mathur2025}. Strong magnetic activity in turn leads to efficient magnetic braking, and at fixed mass and age, metal-rich stars are thus slower rotators than their peers \citep[e.g.][]{Amard2020,Amard2020b,Santos2023}. Therefore, the rotational evolution and the associated timescales depend on metallicity. Praesepe is metal rich \citep[e.g.][]{Cao2023}, while NGC 6811 is close to solar metallicity \citep[e.g.][]{Sandquist2016}. This difference in metallicity might cause the difference in the evolution timescales, with Praesepe evolving faster in terms of rotation.

Core-envelope coupling has been pointed out as the reason for the intermediate-\prot\ gap and stellar spin-down stalling \citep[e.g.][]{McQuillan2014,Curtis2019,Spada2020,Angus2020,Lu2022}. In light of this scenario, \citet{Cao2023} explained the enhanced magnetic activity with the strong internal shear that results from the coupling between the slow envelope and the fast core. The results we found support this hypothesis.
From their models, \citet{Cao2023} concluded that this phase of an enhanced magnetic activity can last from hundreds of millions to billions of years, depending on the mass. As discussed above, the wide range of temperatures for which we found an enhanced activity for NGC 6811 suggests a long-lasting effect, and this agrees with previous results. In addition to the complex stellar evolution, these findings are also significant for the evolution of planetary systems and their environment.

\begin{acknowledgements}
This work was supported by Fundação para a Ciência e a Tecnologia (FCT) through the
research grants UIDB/04434/2020 (DOI: 10.54499/UIDB/04434/2020), UIDP/04434/2020.FCT (DOI: 10.54499/UIDP/04434/2020) and 2022.03993.PTDC (DOI: 10.54499/2022.03993.PTDC). This paper includes data collected by the \kep\, mission and obtained from the MAST data archive at the Space Telescope Science Institute (STScI). Funding for the \kep\ mission is provided by the NASA Science Mission Directorate. STScI is operated by the Association of Universities for Research in Astronomy, Inc., under NASA contract NAS 5–26555. This research was also supported by the International Space Science Institute (ISSI) in Bern, through ISSI International Team project 24-629 (Multi-scale variability in solar and stellar magnetic cycles). We thank the organizers of the 11\textsuperscript{th} Iberian Meeting on Asteroseismology for fostering valuable interactions that contributed to this work. ARGS acknowledges the support from the FCT through the work contract No. 2020.02480.CEECIND/CP1631/CT0001 (DOI: 10.54499/2020.02480.CEECIND/CP1631/CT0001). DGR acknowledges support from the Juan de la Cierva program under contract JDC2022-049054-I. SM\ acknowledges support by the Spanish Ministry of Science and Innovation with the grant no. PID2019-107061GB-C66, and through AEI under the Severo Ochoa Centres of Excellence Programme 2020--2023 (CEX2019-000920-S). S.M. and D.G.R. acknowledge support from the Spanish Ministry of Science and Innovation (MICINN) with the grant No.  PID2023-149439NB-C41. RAG acknowledges the support from the GOLF and PLATO Centre National D'{\'{E}}tudes Spatiales grants. SNB acknowledges support from PLATO ASI-INAF agreement n. 2015-019-R.1-2018. MSC is funded by FCT-MCTES by the contract with reference 2023.09303.CEECIND/CP2839/CT0003 (DOI: 10.54499/2023.09303.CEECIND/CP2839/CT0003). \textit{Software:} Matplotlib \citep{matplotlib}, NumPy \citep{numpy}, Seaborn \citep{seaborn}, pandas \citep{mckinney-proc-scipy-2010_pandas,reback2020pandas}

\end{acknowledgements}

\bibliographystyle{aa}
\bibliography{skumanich}

\begin{appendix}

\section{Location of the intermediate-\prot\ gap}\label{sec:app_gap}

The location of the intermediate-\prot\ gap is not exempt from uncertainties. Furthermore, given the uncertainty on their \prot\ estimate, stars might be misidentified as belonging to the fast-rotating or slowly rotating population. In this section we investigate whether these uncertainties can significantly influence the identification of stars according to their relation to the intermediate-\prot\ gap.

Following the approach in \citet{Santos2023}, the location of the intermediate-\prot\ gap is determined by resorting to the activity-rotation relationship. As shown in Fig.~\ref{fig:sph_prot}, the discontinuity in the stellar evolution leads to a local minimum in the \avsph-\prot\ diagram at a given \teff. We use this local minima as an estimate of the gap location at a given temperature. In the previous sections, the gap location was determined according to the following steps:
\begin{enumerate}
    \item the \kep\ sample is split in \teff\ bins of width 100 K;
    \item for each \teff\ bin, stars are split in log(\prot) bins of with 0.02;
    \item for each log(\prot) bin, the 95\textsuperscript{th} percentile of the \sph\ distribution is computed and taken as an indication of the upper envelope of the activity-rotation relationship;
    \item for each \teff\ bin, taking the upper envelope of the binned \sph-\prot\ relation, we identify the local minimum and corresponding log(\prot);
    \item we fit a third-order polynomial to the log(\prot) as a function of \teff, which we adopt as the location of the intermediate-\prot\ gap (Fig.~\ref{fig:prot_teff}). The resulting polynomial is $\log P_\text{gap}=-1.81\times10^{-10} + 2.59\times10^{-6} T_\text{eff} - 1.24\times10^{-2} T_\text{eff}^2 + 2.11\times10^{1} T_\text{eff}^3$.
\end{enumerate}

\begin{figure}[b]
    \centering
    \includegraphics[width=\hsize]{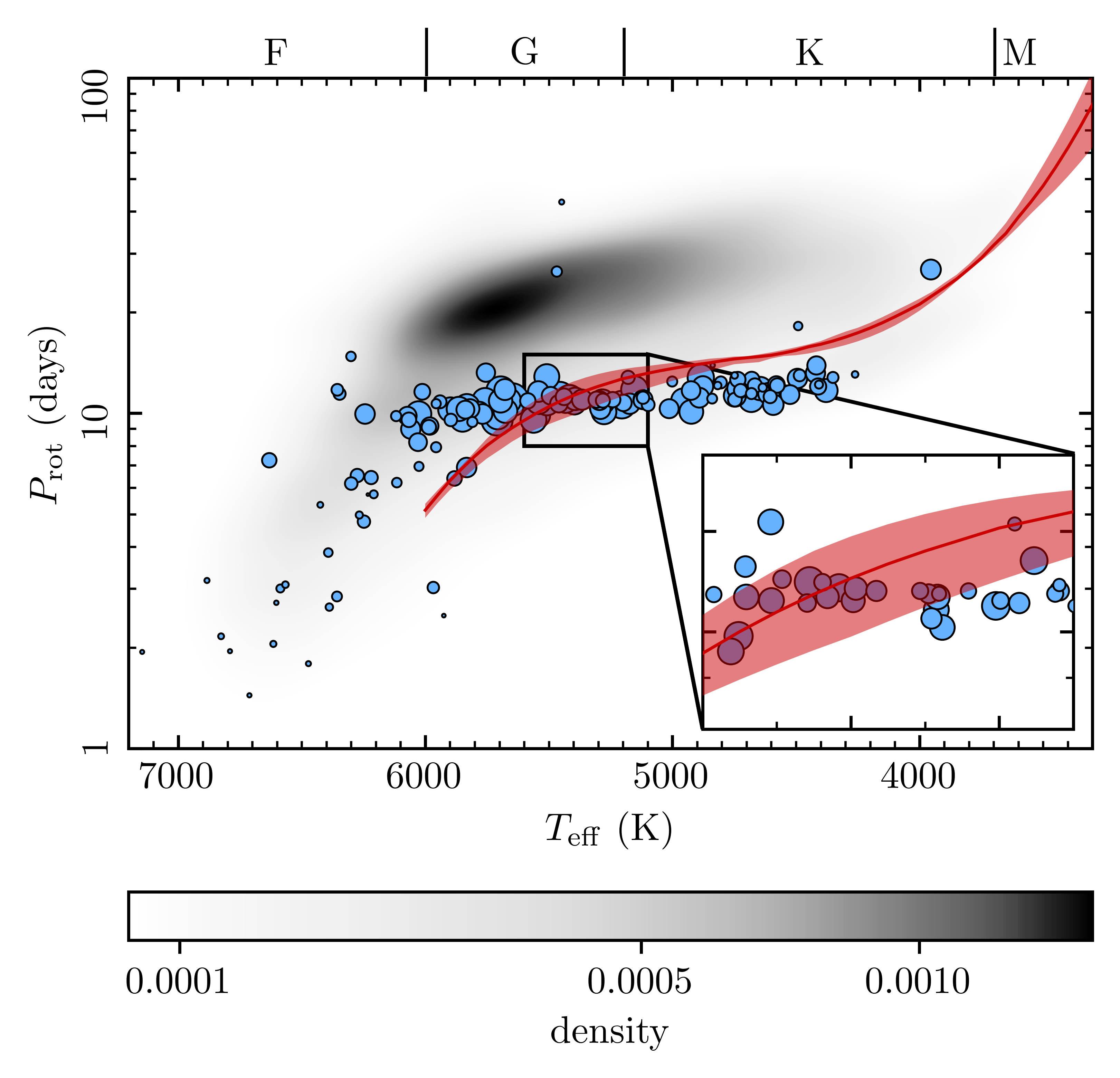}
    \caption{Same as in Fig.~\ref{fig:prot_teff}. The reference  \kep\ sample is now shown in grayscale. The red line and shaded region summarize the results of bootstrapping where the randomly varied \teff, \prot, and the size of the respective bins. The solid line indicates the median of 1000 repetitions, while the shaded region indicates the $1\sigma$.}
    \label{fig:gap_unc}
\end{figure}

In this section we perform a bootstrapping to account for the potential sources of uncertainty associated with the determination of the intermediate-\prot\ gap. The average uncertainties on \teff\ and \prot\ are $\sim102$ K ($\sim 1.9\%$) and $\sim 1.92$ days ($\sim 9.8\%$), respectively, for the reference \kep\ sample. We randomly vary the size of the \teff\ and log(\prot) bins in points 1 and 2: \teff\ bins can have a width of 50, 100, 150, and 200 K; log(\prot) bins can have a width of 0.015, 0.020, 0.025, and 0.030. In each iteration, the \teff\ and \prot\ of each star in the reference \kep\ sample are taken randomly from a Gaussian distribution taking the literature values for the mean and uncertainty of each parameter. In the case of \teff, as we are assuming a symmetric distribution for this exercise, we take the average uncertainty (both 16\textsuperscript{th} and 84\textsuperscript{th} percentiles are available in the literature, i.e. in \citealt{Berger2020} and \citealt{Mathur2017}). We repeat this process for 1000 iterations, performing the five steps listed above accordingly. 

\begin{figure}[h]
    \centering
    \includegraphics[width=\hsize]{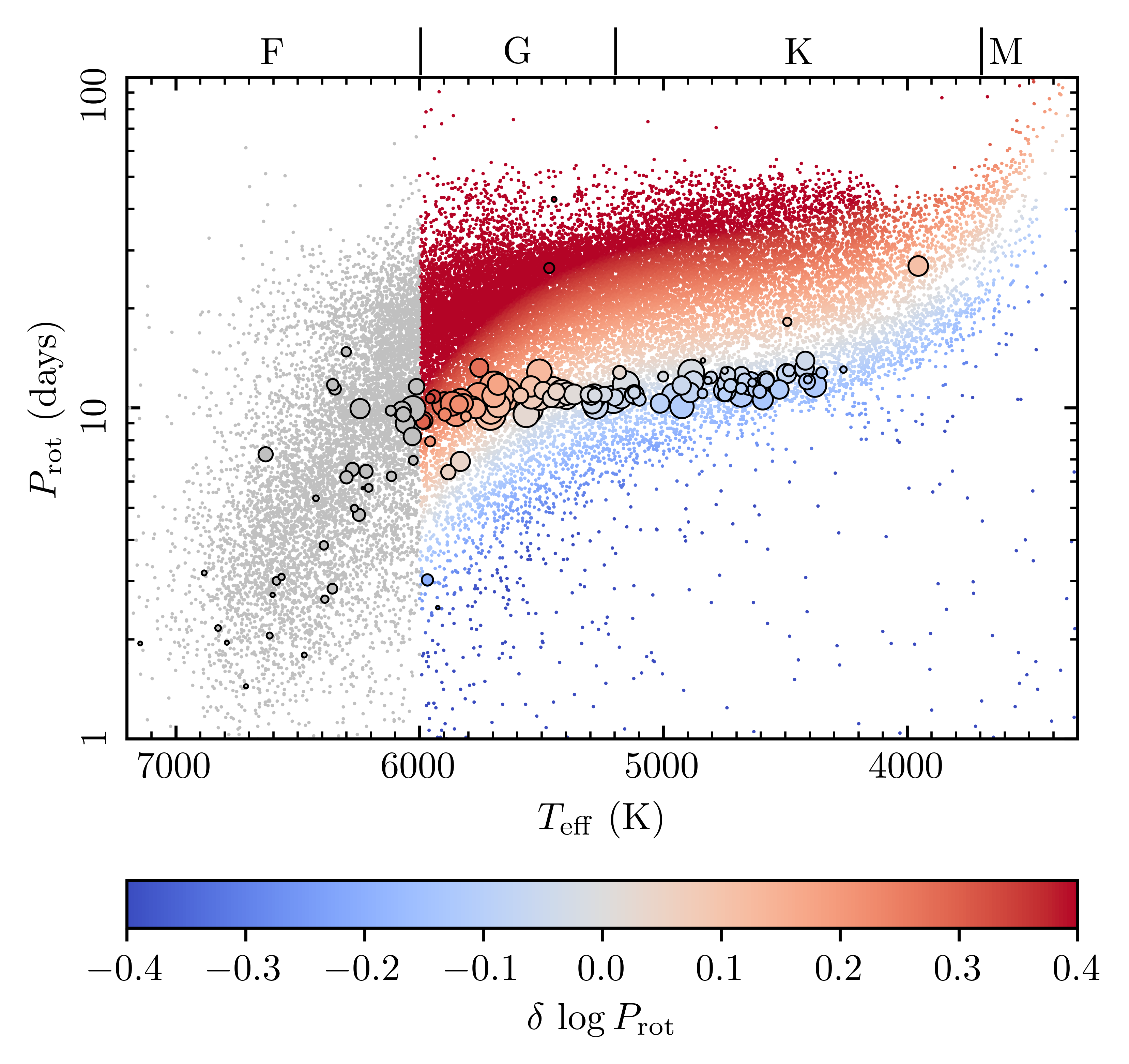}
    \caption{Same as in the main panel of Fig.~\ref{fig:prot_teff} but the color code indicates the distance to the intermediate-\prot\ gap for both \kep\ single MS stars (dots) and NGC~6811 (outlined circles). F stars, which do not show evidence of \prot\ bimodality, are depicted in gray.}
    \label{fig:prot_teff_gapcolor}
\end{figure}

Figure~\ref{fig:gap_unc} and Table~\ref{tab:gap} summarize the results from the bootstrapping. The solid red line shows the median gap location, while the red-shaded region indicates the 16\textsuperscript{th} and 84\textsuperscript{th} percentiles (1$\sigma$). The location of the gap is more uncertain towards the lower edge of the parameter space, which could be related to the small sample size at low \teff. Furthermore, around 3500 K, where the transition from partial to fully convective occurs, the gap disappears \citep{Lu2022}. These two effects contribute to the large uncertainty of the low-temperature gap location. Nevertheless, the uncertainty associated with the gap location remains relatively small. The NGC 6811 crossing occurs around 5400 K according to this exercise, not changing the conclusions and discussion of this work.
As a reference to Figs.~\ref{fig:sph_prot} and \ref{fig:sph-teff-kep}, as well as figures in the next section, Fig.~\ref{fig:prot_teff_gapcolor} exemplifies the color code according to distance to the intermediate-\prot\ gap in Sect.~\ref{sec:prot}.

\begin{table}[h]
    \centering
    \caption{Intermediate-\prot\ gap.\label{tab:gap}}
    \begin{tabular}{cccc}
\teff (K) & $P_{\text{gap},16^\text{th}}$ (days)& $P_{\text{gap},50^\text{th}}$ (days)& $P_{\text{gap},84^\text{th}}$ (days) \vspace{0.05cm}\\\hline
3200 & 77.30 & 117.15 & 160.87\\ 
3250 & 69.09 & 98.34 & 128.36\\ 
3300 & 62.07 & 83.43 & 105.00\\ 
3350 & 56.06 & 71.51 & 87.89\\ 
3400 & 50.88 & 61.90 & 74.40\\ 
3450 & 46.47 & 54.09 & 63.65\\ 
3500 & 42.75 & 47.70 & 55.03\\ 
3550 & 39.35 & 42.65 & 47.77\\ 
3600 & 36.02 & 38.48 & 41.80\\ 
3650 & 33.16 & 34.45 & 37.02\\ 
3700 & 30.59 & 31.73 & 33.38\\ 
3750 & 28.55 & 29.05 & 30.35\\ 
3800 & 26.64 & 26.98 & 27.89\\ 
3850 & 24.78 & 25.14 & 25.90\\ 
3900 & 23.00 & 23.71 & 24.48\\ 
3950 & 21.52 & 22.39 & 23.03\\ 
4000 & 20.24 & 21.15 & 21.97\\ 
4050 & 19.17 & 20.24 & 21.04\\ 
4100 & 18.27 & 19.40 & 20.18\\ 
4150 & 17.53 & 18.63 & 19.36\\ 
4200 & 16.91 & 17.96 & 18.62\\ 
4250 & 16.40 & 17.38 & 17.99\\ 
4300 & 15.98 & 16.87 & 17.52\\ 
4350 & 15.63 & 16.43 & 16.97\\ 
4400 & 15.36 & 16.06 & 16.42\\ 
4450 & 15.08 & 15.78 & 16.05\\ 
4500 & 14.90 & 15.41 & 15.74\\ 
4550 & 14.82 & 15.13 & 15.46\\ 
4600 & 14.57 & 14.90 & 15.20\\ 
4650 & 14.21 & 14.76 & 14.99\\ 
4700 & 14.11 & 14.64 & 14.78\\ 
4750 & 14.01 & 14.52 & 14.76\\ 
4800 & 13.70 & 14.34 & 14.64\\ 
4850 & 13.39 & 14.14 & 14.55\\ 
4900 & 13.09 & 13.98 & 14.38\\ 
4950 & 12.79 & 13.80 & 14.25\\ 
5000 & 12.49 & 13.60 & 14.13\\ 
5050 & 12.19 & 13.39 & 13.99\\ 
5100 & 11.88 & 13.16 & 13.83\\ 
5150 & 11.57 & 12.92 & 13.71\\ 
5200 & 11.25 & 12.68 & 13.55\\ 
5250 & 10.92 & 12.35 & 13.35\\ 
5300 & 10.58 & 12.03 & 13.09\\ 
5350 & 10.24 & 11.68 & 12.79\\ 
5400 & 9.89 & 11.30 & 12.44\\ 
5450 & 9.59 & 10.90 & 12.03\\ 
5500 & 9.28 & 10.46 & 11.54\\ 
5550 & 8.96 & 10.00 & 10.99\\ 
5600 & 8.64 & 9.52 & 10.39\\ 
5650 & 8.22 & 9.03 & 9.76\\ 
5700 & 7.76 & 8.53 & 9.10\\ 
5750 & 7.33 & 8.02 & 8.41\\ 
5800 & 6.85 & 7.44 & 7.67\\ 
5850 & 6.35 & 6.86 & 7.01\\ 
5900 & 5.90 & 6.29 & 6.42\\ 
5950 & 5.40 & 5.69 & 5.89\\ 
6000 & 4.88 & 5.14 & 5.37\\   \hline
    \end{tabular}
    \tablefoot{Intermediate-\prot\ gap (median) and respective $1\sigma$ values resulting from the bootstrapping exercise. The \teff\ resolution was set at 50 K. The subscripts (16\textsuperscript{th}, 50\textsuperscript{th}, 84\textsuperscript{th}) indicate the percentile corresponding to each column.}
\end{table}


\section{Magnetic-activity variation}\label{sec:app_Sphvariation}

\begin{figure}[h]
    \centering
    \includegraphics[width=\hsize]{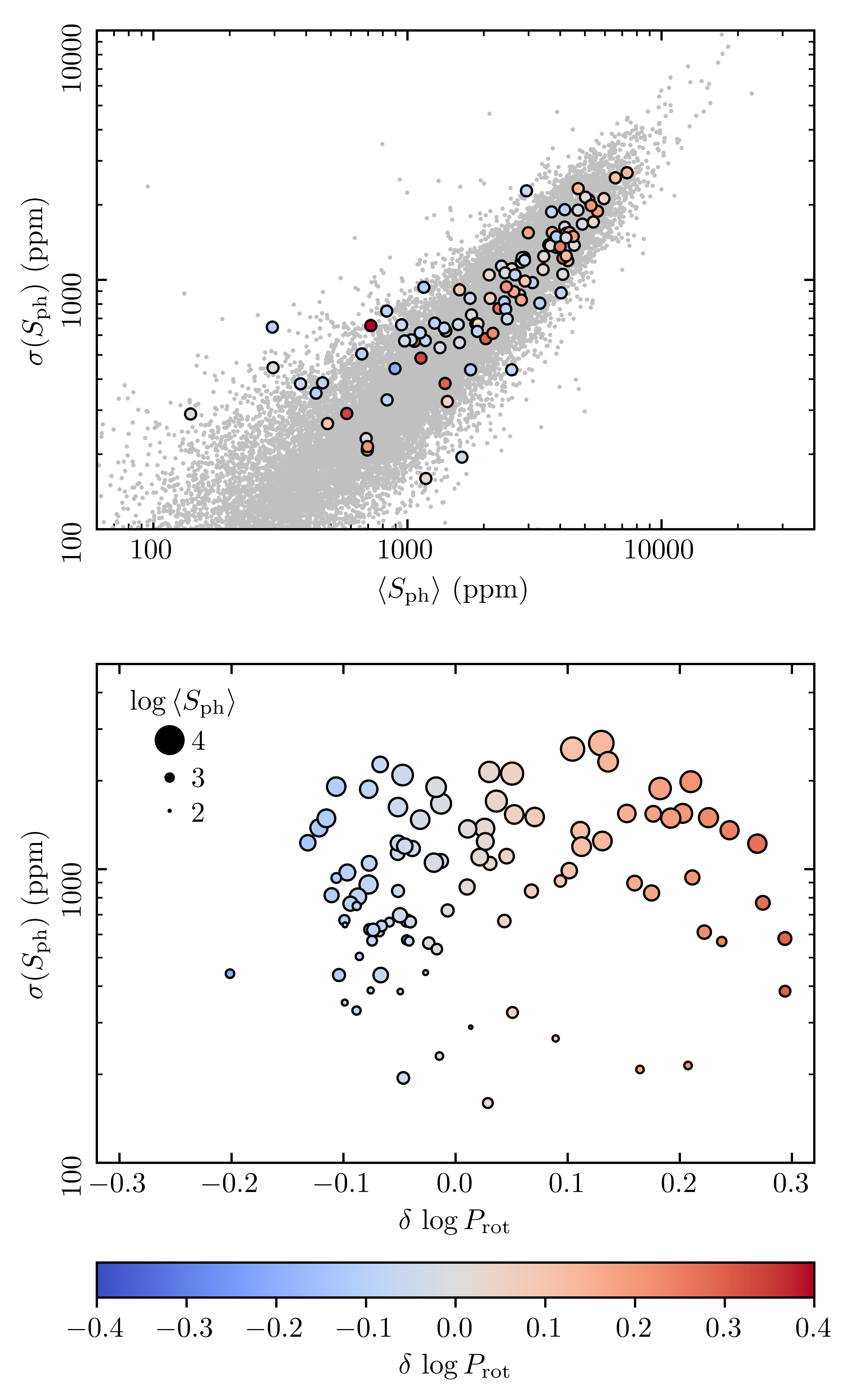}
    \caption{{\it Top:} \stdsph\ as a function of \avsph\ for the NGC~6811 members. The F-type stars are not included in this representation as there is no evidence for \prot\ bimodality in this regime. The reference \kep\ sample is shown in gray, where the \stdsph\ values were taken from \citet{Santos2023}. {\it Bottom:} \stdsph\ as a function of the distance to the intermediate-\prot\ gap, where a few data points with large $|\delta\log P_\text{rot}|$ were omitted for illustration purposes. The symbol size indicates the average magnetic activity level. The color code in both panels corresponds to the distance to the intermediate-\prot\ gap.}
    \label{fig:stdsph}
\end{figure}

Magnetic activity varies over time and it has been shown that stars with stronger activity are those with more significant temporal variations \citep[e.g.][]{Radick1998,EBrown2022,Santos2023}. Similarly to what was found for the reference \kep\ sample, there is a tight relation between \avsph\ and \stdsph\ for the NGC~6811 members (top panel of Fig.~\ref{fig:stdsph}). For guidance, from the linear regression applied to the \kep\ sample (in gray), we obtain $\log\sigma(S_\text{ph})=(0.842\pm0.002)\log\langle S_\text{ph}\rangle+(0.076\pm0.007)$ \citep[see][for a more detailed analysis]{Santos2023}.

As the stars near the intermediate-\prot\ gap are among those with the largest \avsph, these also exhibit significant variations in the magnetic-activity level. However, there is no clear relation between the \stdsph\ and the distance to the gap (bottom panel of Fig.~\ref{fig:stdsph}). This indicates that the stars with enhanced magnetic activity do not suffer from a particular bias related to the \sph\ variability over the 4-year observations.

\begin{figure}[h]
    \centering
    \includegraphics[width=\hsize]{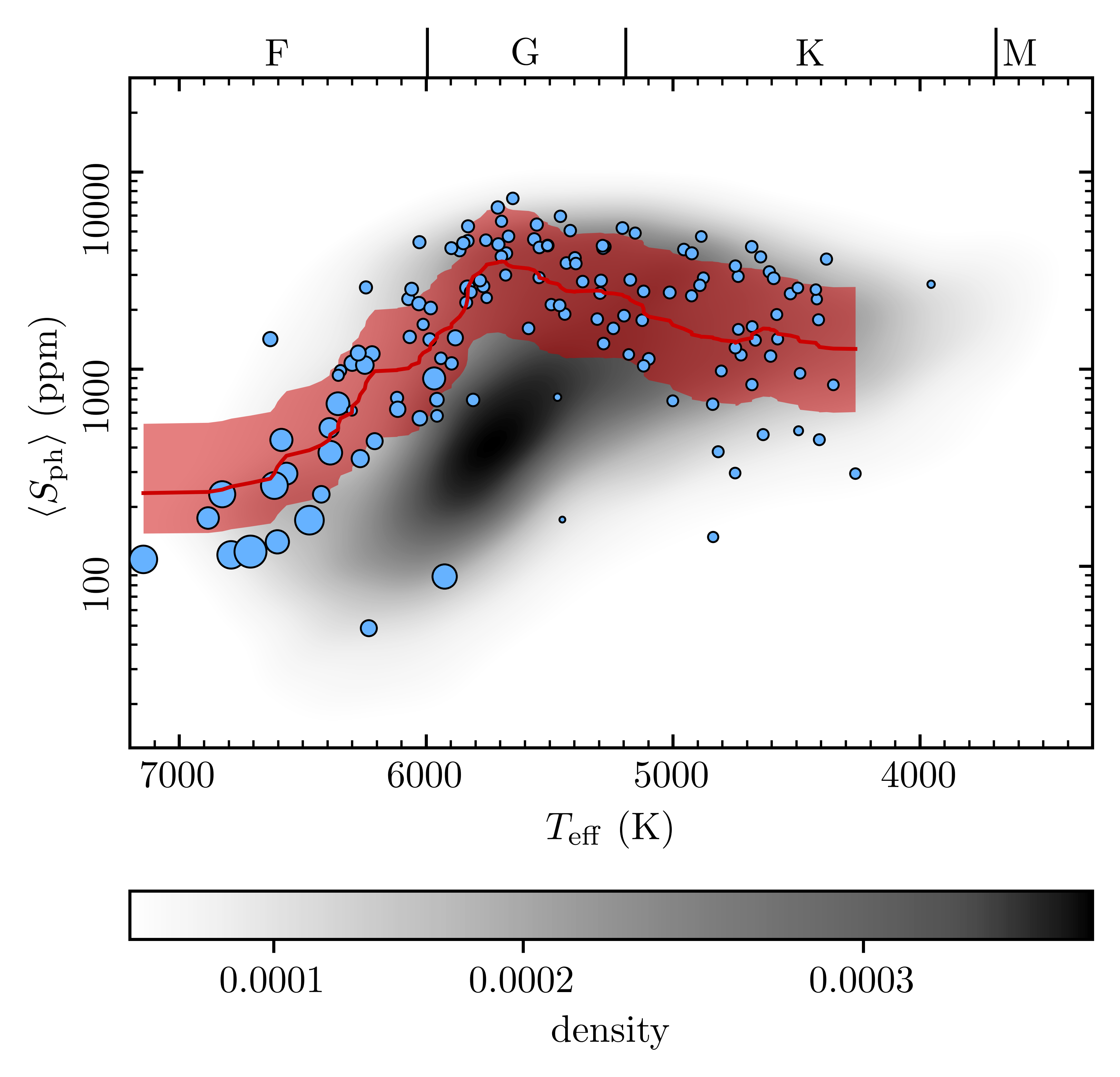}
    \caption{Same as in Fig.~\ref{fig:sph_teff}. The reference  \kep\ sample is now shown in grayscale. The solid red line shows the same as the gray line in Fig.~\ref{fig:sph_teff}, while the shaded region indicates the range of \sph\ values over the 4-yr \kep\ observations.}
    \label{fig:sigsph_teff}
\end{figure}

Figure \ref{fig:sigsph_teff} is similar to Fig.~\ref{fig:sph_teff} but adds the shaded region indicating the range of \sph\ values over the \kep\ observations. In particular, the lower and upper edges correspond to a uniform filter (smoothing) applied to minimum and maximum \sph\ for the cluster members. The shaded region and the respective edges show the same behavior as the average \sph\ (solid line; already shown in Fig.~\ref{fig:sph_teff}). This confirms that no bias is introduced by taking the \avsph\ to characterize the stars in this sample.

\section{Observational bias at low activity?}\label{sec:low_activity}

We identified an overdensity and overlap of activity sequences for stars significantly slower than the intermediate-\prot\ gap (Sect.~\ref{sec:kepler}). This corresponds to stars close to the upper edge of the rotation distribution (slow rotators). This rather sharp edge is a consequence of two potential contributions:
\begin{itemize}
    \item detection limit: slow rotators are typically weakly active and, thus, characterized by small-amplitude rotation modulations, which can be easily hidden within the noise \citep[e.g.][]{Masuda2022,Masuda2022c};

    \item weakened magnetic braking (WMB): due to a loss of efficiency in the magnetic braking, possibly arising from a switch from large-scale to small-scale magnetic fields \citep{Metcalfe2017,Metcalfe2022}, \prot\ is predicted to evolve slowly in this region of the parameter space \citep[e.g.][]{vanSaders2016,vanSaders2019}. Indeed pile-ups of stars near the upper edge in \prot\ have been identified \citep{Hall2021,David2022,Metcalfe2023}.
\end{itemize}

In the WMB scenario, while \prot\ is expected to not change significantly after reaching the critical Ro, the magnetic activity would continue its gradual evolution along the MS \citep[e.g.][]{Lorenzo-Oliveira2018}. For that reason, the pile-up of activity sequences at low activity was unexpected. 

In this section we explore the possibility of observational bias leading to this activity pile-up. To that end, we investigate the properties of the \prot\ and \sph\ distributions as a function of the \kep\ magnitude (Kp). The fainter the object, the noisier the light curve. Therefore, we expect to be limited to the detection of high-amplitude signals for large Kp (faint stars), while for brighter stars, weaker signals are expected to be detectable.

\begin{figure*}
    \centering
    \includegraphics[width=\hsize]{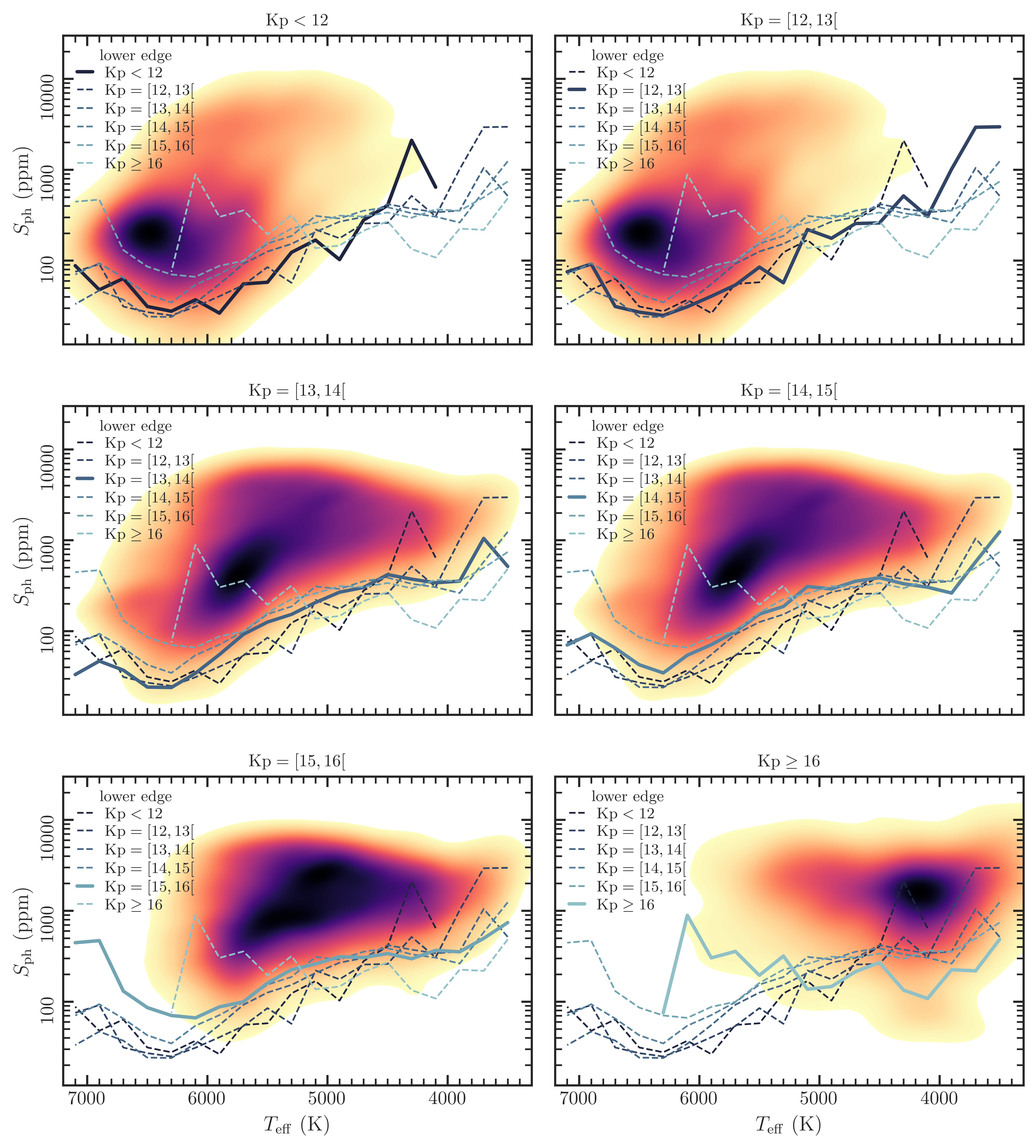}
    \caption{Density map for \sph\ as a function of \teff, where darker colors correspond to higher density (consistent with the color code in Fig.~\ref{fig:sph_teff}). Each panel corresponds to a different \kep\ magnitude interval: from bright (top left) to faint (bottom right). The lines represent the lower edge (5\textsuperscript{th} percentile) of the \sph\ distribution: solid line -- the same magnitude of the density plot; dashed lines -- other magnitude intervals, for reference. The color code of the lines also indicates the respective Kp.}
    \label{fig:sph_kp}
\end{figure*}

We split the reference \kep\ sample according to Kp. Figure~\ref{fig:sph_kp} shows the respective \sph-\teff\ diagrams. Some anticipated trends are evident: a small number of bright lower-mass (lower-\teff) stars are observed, while the reverse holds true for higher-mass (higher-\teff) stars. In each panel, the lower edge (5\textsuperscript{th} percentile) of the \sph\ distribution is represented by a solid line. For easier comparison, the lower edges found for the remainder Kp intervals are shown by the dashed lines. The line color indicates the respective Kp interval. The bias described above can be identified, with few small-\sph\ targets found beyond magnitude 16. From brighter to fainter targets, there is a gradual shift of the lower \sph\ edge towards larger values. However, this trend is more noticeable at higher temperatures, while at lower temperatures the behavior seems to invert. 

\begin{figure*}[ht]
    \centering
    \includegraphics[width=0.053\hsize,trim=10 0 745 0mm,clip]{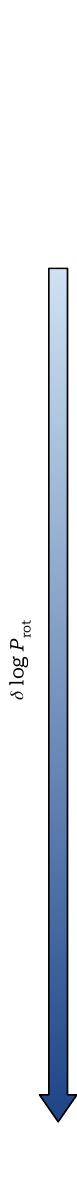}
    \includegraphics[width=0.469\hsize]{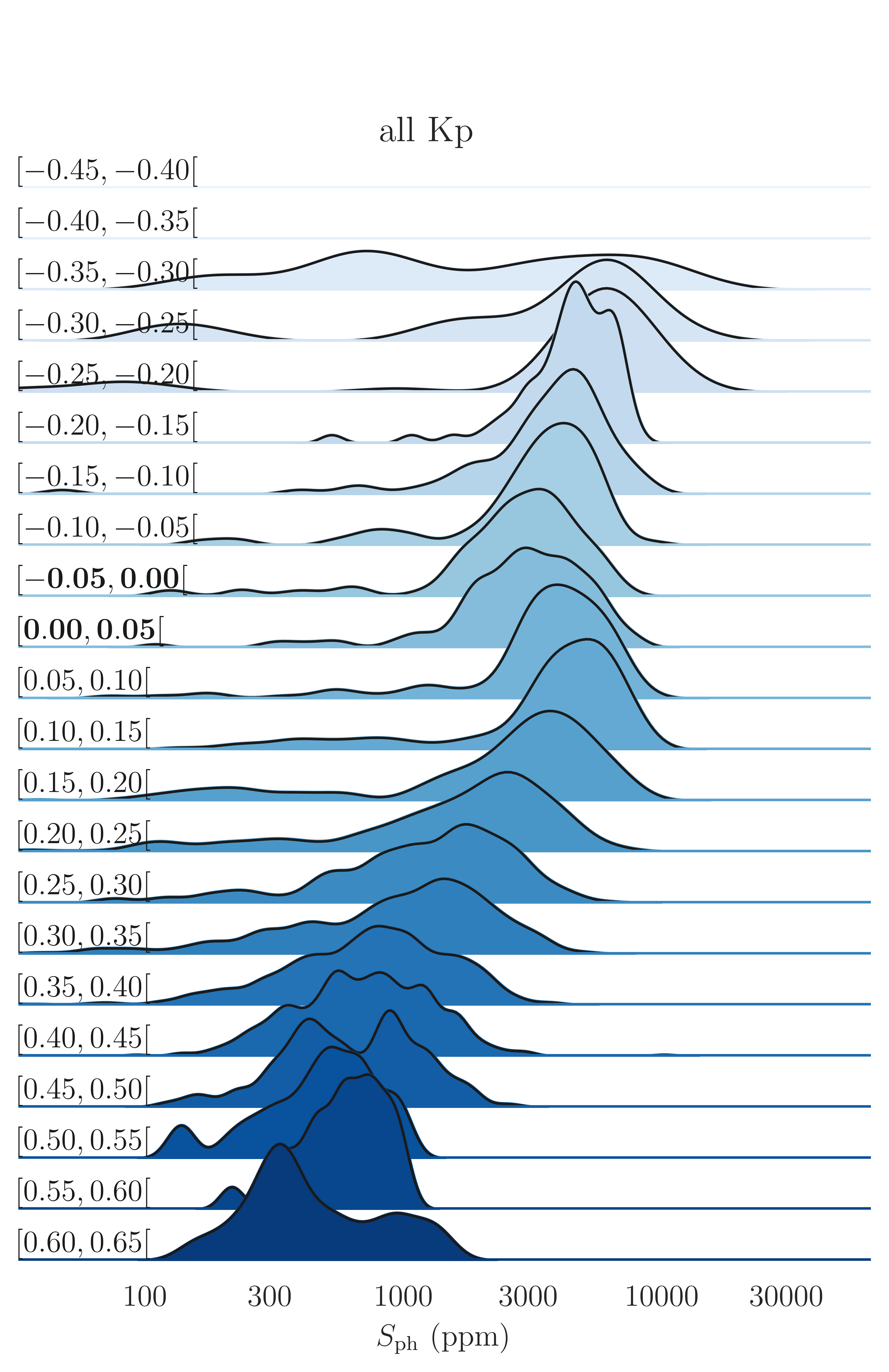}
    \includegraphics[width=0.469\hsize]{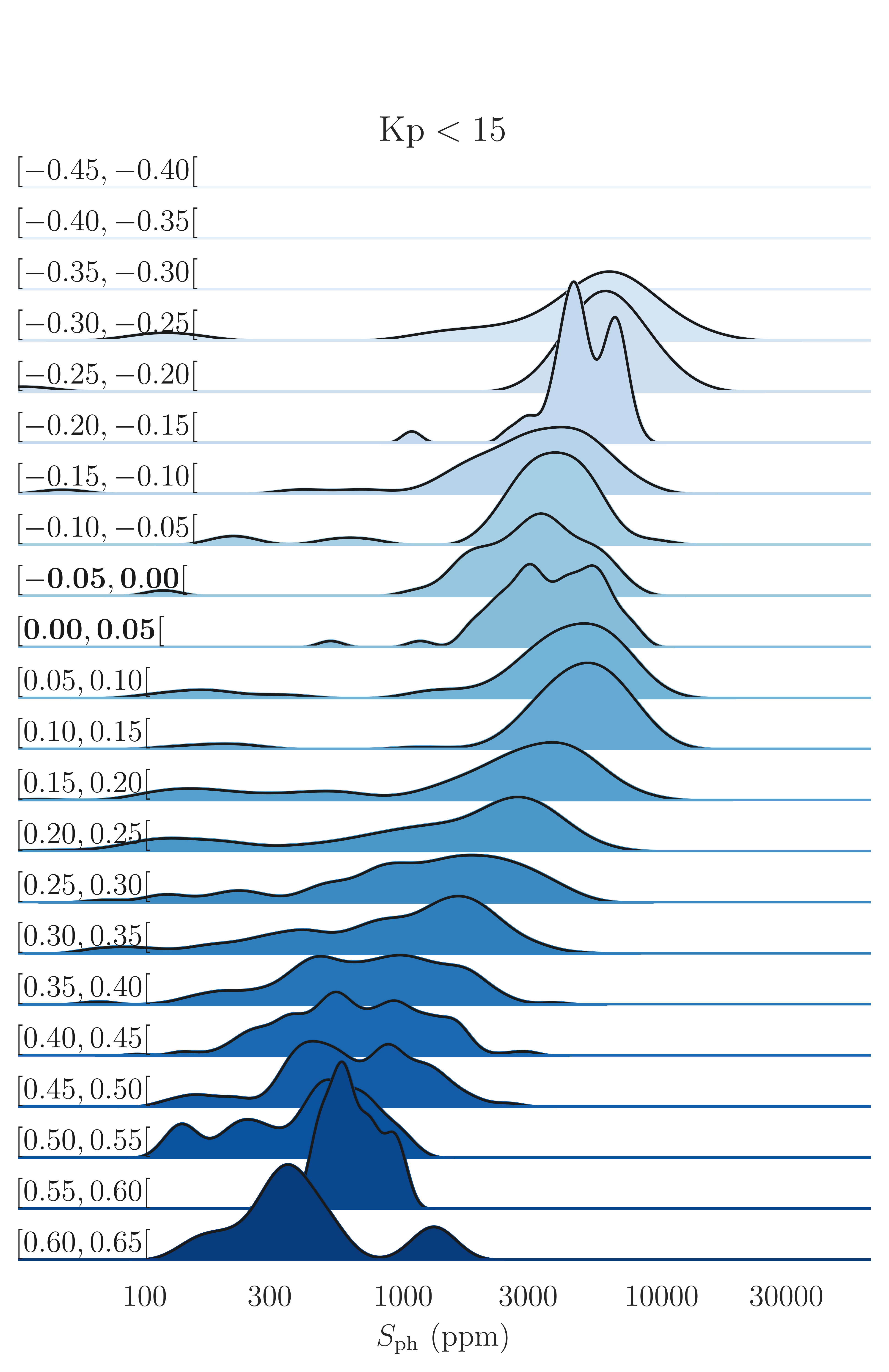}
    \caption{\sph\ distribution for \teff\ within 5200 and 5400 K for different rotation regimes. Rows and colors represent different distances to the intermediate-\prot\ gap ($\delta \log P_\text{rot}$; annotated on each panel). The closest rows to zero $\delta \log P_\text{rot}$ are marked in boldface for the easiest identification. The bins are originally set between -0.80 and 0.80 with steps of 0.05 (left and right panels of Figs.~\ref{fig:5200_sph_kp_deltaP} and \ref{fig:5500_sph_kp_deltaP}), but only bins with more than five stars are considered for representation. The exception is the bins corresponding to $-0.45\leq\delta \log P_\text{rot}<-0.30$ on the right, which are kept for consistency. The left panels correspond to all \kep\ magnitudes, while the right panels to Kp $<15$.}
    \label{fig:5200_sph_kp_deltaP}
\end{figure*}

\begin{figure*}[h!]
    \centering
    \includegraphics[width=0.053\hsize,trim=10 0 745 0mm,clip]{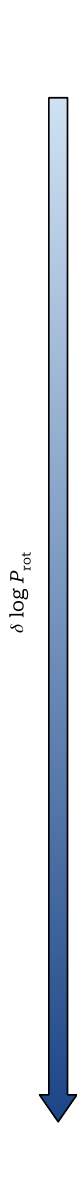}
    \includegraphics[width=0.469\hsize]{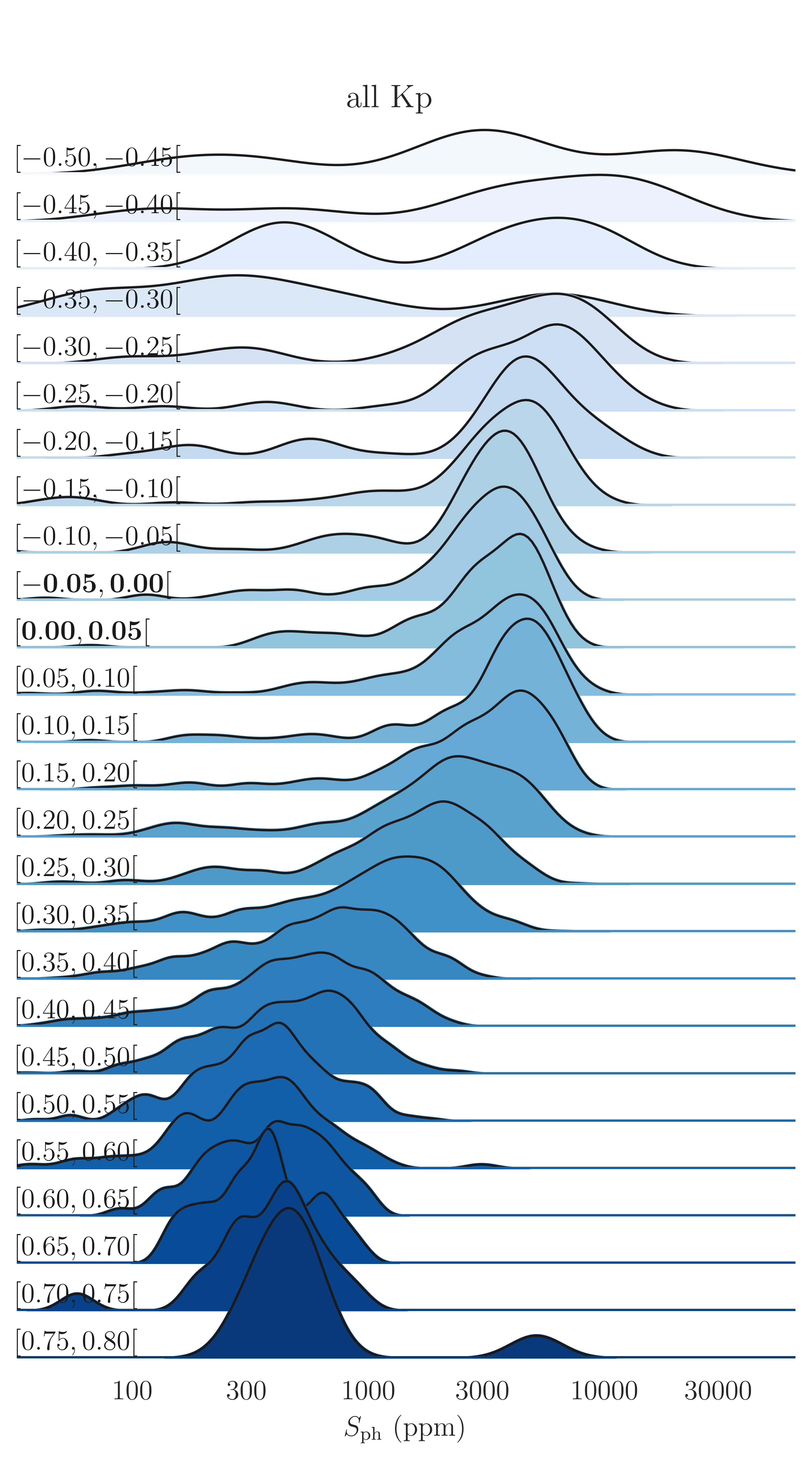}
    \includegraphics[width=0.469\hsize]{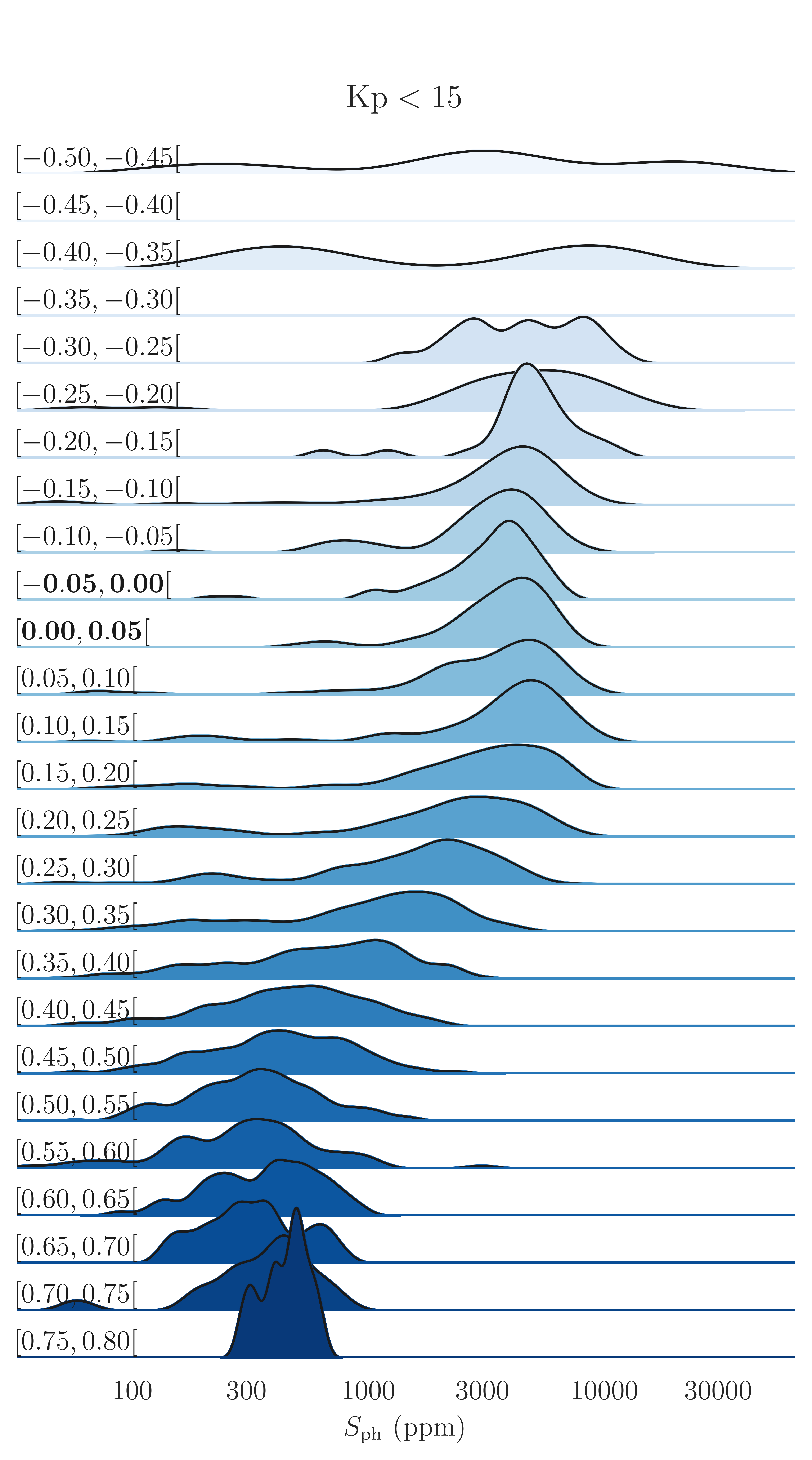}
    \caption{Same as in Fig.~\ref{fig:5200_sph_kp_deltaP} but for \teff\ within 5500 and 5700 K.} 
    \label{fig:5500_sph_kp_deltaP}
\end{figure*}

If this observational bias resulted in a pile-up of activity sequences, in addition to restricting the ability to detect low-amplitude signals, the noise would also contribute to the inferred \sph\ value. We remind that \sph\ is already corrected for the photon-shot noise \citep[e.g.][]{Santos2021ApJS}. In this regime, the amplitude of the brightness variations due to active regions and due to noise would be comparable. If this was the sole cause for the activity sequence pile-up, this pile-up would not be expected at brighter magnitudes. However, when limiting the sample to magnitudes smaller than 15, the pile-up is still observed (Figs.~\ref{fig:5200_sph_kp_deltaP} and \ref{fig:5500_sph_kp_deltaP}). Furthermore, we would not expect \sph\ to increase towards longer \prot, which is observed at high \teff. This leads to the conclusion that, in addition to the potential observational bias, the pile-up of activity sequences for slow rotators is also of physical origin.

Figures \ref{fig:5200_sph_kp_deltaP} and \ref{fig:5500_sph_kp_deltaP} show the \sph\ distribution for \teff\ within 5200-5400 K and 5500-5700 K, respectively. The left panels include all targets, while the right panels only include those with $\text{Kp}<15$. The rows and colors indicate the distance to the intermediate-\prot\ gap, with light blue corresponding to \prot\ shorter than the gap and dark blue to \prot\ longer than the gap. These are similar to Fig.~\ref{fig:sph_prot}, but better depict the distribution at a fixed distance to the gap. The rows with stars near the gap ($\delta\log P_\text{rot}\sim0$) are highlighted in boldface. For cooler stars (Fig.~\ref{fig:5200_sph_kp_deltaP}), three near-plateaux (overlapping of activity sequences) can be identified:
\begin{enumerate}
    \item $P_\text{rot}\ll$ intermediate-\prot\ gap: saturated regime, which is not well probed by \kep\ \citep[e.g.][]{Santos2024} and where the sample sizes are small;

    \item $P_\text{rot}\approx$ intermediate-\prot\ gap: more than the stagnation of the activity evolution, as described above, the activity sequences near this transition cross each other as stars immediately after the gap exhibit enhanced activity;

    \item $P_\text{rot}\gg$ intermediate-\prot\ gap: upper edge of the \prot\ distribution and lower edge of the \sph\ distribution. 
\end{enumerate}
For hotter stars (Fig.~\ref{fig:5500_sph_kp_deltaP}), the first plateau is not very well identified, which was expected given the known \sph\ distribution for G dwarfs \citep[e.g.][]{Santos2021ApJS}. Nevertheless, the other two plateaux are well-identified. At this \teff\ regime, \sph\ is seen to increase at large Ro \citep{Mathur2025}, which is consistent with the behavior seen here at long \prot.

\begin{figure*}
    \centering
    \includegraphics[width=\hsize]{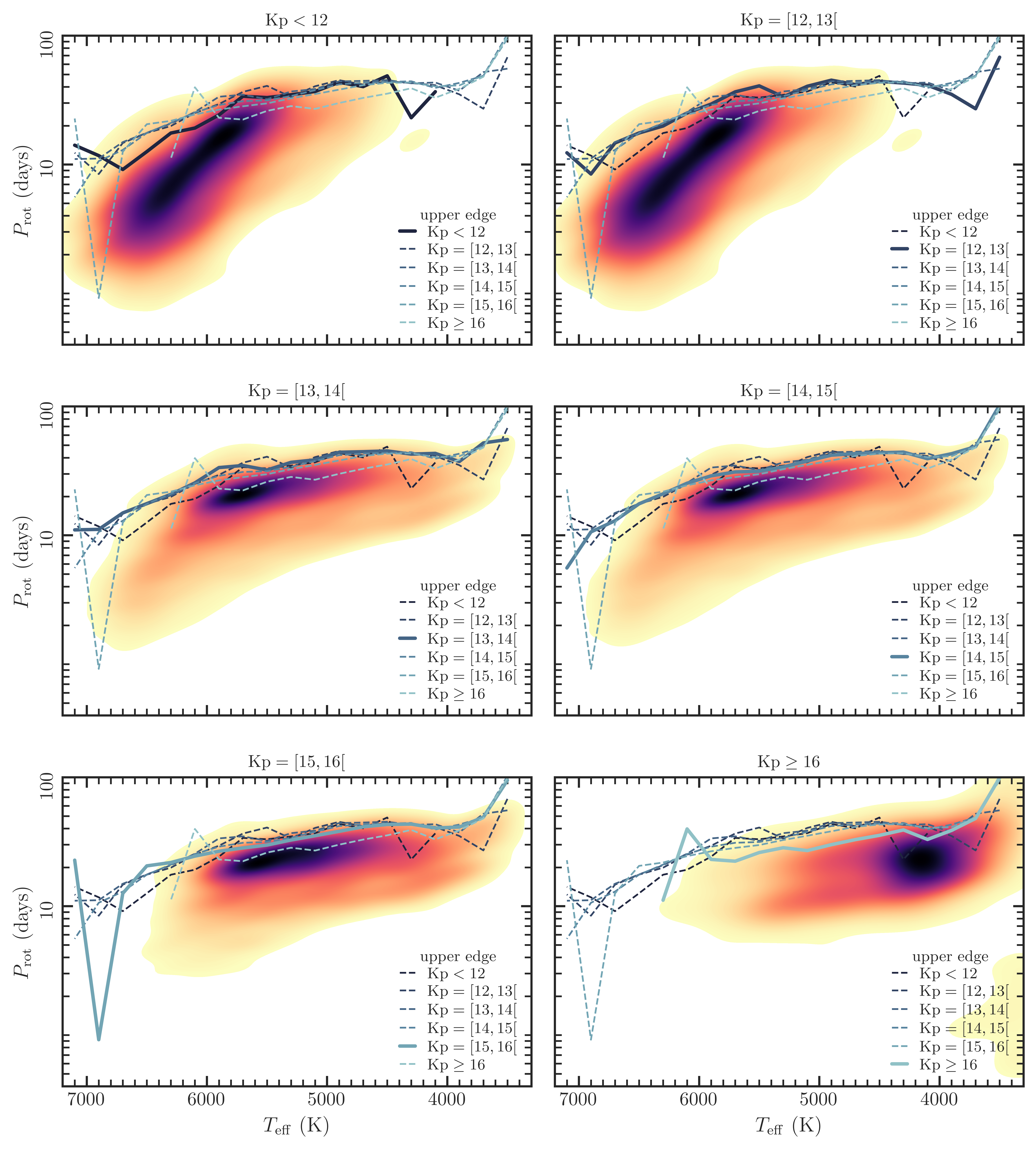}
    \caption{Same as in Fig.~\ref{fig:sph_kp} but for the rotation period and the respective upper edge (95\textsuperscript{th} percentile). }
    \label{fig:prot_kp}
\end{figure*}

Figure~\ref{fig:prot_kp} shows the same as Fig.~\ref{fig:sph_kp} but for \prot\ and the respective upper edge (95\textsuperscript{th} percentile) of the distribution. In this case, the bias on \prot\ is less noticeable than that for \sph; however, the upper edge for stars with $\text{Kp}>16$ is located at periods significantly shorter than those of brighter stars. This supports the hypothesis that we lose the ability to detect the signals of faint slow rotators.

\end{appendix}

\end{document}